%% file: ms.tex
\documentclass{emulateapj}

\newcommand{\eg}{{e.g. \/}}
\newcommand{\ie}{{i.e. \/}}
\newcommand{\etc}{{etc.}}
\newcommand{\etal}{{et al. \/}}
\newcommand{\bump}{1.6$\mu$m bump}
\newcommand{\um}{$\mu$m}
\newcommand{\Msun}{\ensuremath{{\rm M}_{\odot}}}

\newcommand{\Zsun}{\ensuremath{{\rm Z}_{\odot}}}

\slugcomment{ApJ in press}

\shorttitle{The 1.6$\mu$m bump and galaxies at redshift 2}
\shortauthors{Sorba \& Sawicki}

\begin{document}


\title{Using the 1.6$\mu$\lowercase{m} Bump to Study Rest-frame NIR Selected Galaxies at Redshift 2}


\author{Robert Sorba and Marcin Sawicki}
\affil{ Department of Astronomy and Physics, Saint Mary's University, 923 Robie Street, Halifax,
Nova Scotia, B3H 3C3, Canada }
\email{sawicki@ap.smu.ca}




\begin{abstract}
We explore the feasibility and limitations of using the \bump\ as a photometric redshift indicator and selection technique and use it to study the rest-frame $H$-band galaxy luminosity and stellar mass functions at redshift $z$$\sim$2. We use publicly available Spitzer/IRAC images in the GOODS fields and find that color selection in the IRAC bandpasses alone is comparable in completeness and contamination to $BzK$ selection. We find that the shape of the \bump\ is robust, and photometric redshifts are not greatly affected by choice of model parameters. Comparison with spectroscopic redshifts shows photometric redshifts to be reliable. We create a rest-frame NIR selected catalog of galaxies at $z\sim2$ and construct a galaxy stellar mass function (SMF). Comparisons with other SMFs at approximately the same redshift but determined using shorter wavelengths show good agreement. This agreement suggests that selection at bluer wavelengths does not miss a significant amount of stellar mass in passive galaxies. Comparison with SMFs at other redshifts shows evidence for the downsizing scenario of galaxy evolution. We conclude by pointing out the potential for using the \bump\ technique to select high-redshift galaxies with the JWST, whose $\lambda > 0.6 \mu$m coverage will not be well suited to selecting galaxies using techniques that require imaging at shorter wavelengths.
\end{abstract}





\section{Introduction}
\label{intro}

\subsection{Background}

In the last decade, large photometric and spectroscopic galaxy surveys carried out at numerous wavelengths have greatly increased our knowledge about the evolution of galaxies over time. The evolution of galaxies in the universe is represented in the star-formation rate (SFR) density plot (Lilly \etal 1996, Madau \etal 1996, Sawicki, Lin \& Yee 1997), which is a diagram displaying the star formation rate density (usually in units of \Msun/yr/Mpc$^3$) of galaxies as a function of redshift (see Hopkins 2004 for a summary). The SFR density is seen to have a plateau from $z\sim3-2$ that declines sharply at lower redshifts. However, because the SFR is an instantaneous parameter, it has limitations for studying the evolution of galaxies. The stellar mass, which is linked to the entire star formation history of a galaxy, would be a much more appropriate parameter to study galaxy evolution. 

There is increasing evidence showing that the evolution of galaxies follows a ``downsizing" scenario (Cowie \etal 1996), where the most massive galaxies end their star formation first, and star formation shifts to less massive galaxies at more recent times (Heavens \etal 2004, Juneau \etal 2005, Bundy \etal 2006, Tresse \etal 2007). These observations are in contrast to simple interpretations of hierarchical theory, which suggest that small galaxies should form first and larger ones later. Many classical models of galaxy evolution assuming a cold dark matter (CDM) universe predict that the most massive galaxies are created at later times through the merger of smaller halos (\eg Kauffmann \etal 1993, Baugh \etal 1998, De Lucia \etal 2006). Observations detailing how stellar mass evolves with time are essential in order to attempt to resolve this discrepancy.

Infrared (IR) observations are well suited to the study of stellar mass. The flux at rest-frame near-infrared (NIR) wavelengths comes predominantly from relatively older, cooler, less massive stars, which are where the majority of stellar mass in galaxies lies. NIR fluxes are also relatively immune to reddening effects due to extinction from dust, and it is therefore relatively straightforward to derive stellar masses from the NIR flux. This is in contrast to observations at bluer wavelengths, where the flux arises from short-lived, massive stars that contribute little to the total stellar mass. Moreover, stellar mass determinations from bluer observations are non-trivial, as great care must be taken to account for extinction effects, plus the true mass of the galaxy must be inferred by assuming an initial mass function (IMF) to determine the ratio of more massive stars to less massive ones.

In addition, bluer wavelengths may miss a large population of passive galaxies that will be very faint at UV/optical wavelengths, but still bright in NIR and IR. For example, Lyman Break selection (Steidel \etal 1999) is sensitive only to star-forming galaxies, and thus may be biased against quiescent (``red and dead") galaxies. This passive galaxy population may contribute significantly to the stellar mass density of the universe, and could even dominate at some redshifts.

 \subsection{Photometric Redshifts}
The redshift ($z$) of a galaxy can be used as a substitute measure of its distance, or --- similarly --- its lookback time. At $z=1$, the universe was approximately 7 Gyr old, or half its present age. At $z=2$, it was $\sim$ 4 Gyr old, and only $\sim$3 Gyr old at $z=3$. Determining the redshifts to galaxies is often a necessary part of cosmological studies, and is very important when studying how populations of galaxies change over time.
  
Traditionally, redshifts are determined spectroscopically by measuring the shift in the central wavelength of specific emission or absorption lines. However, for large samples of faint objects the spectroscopic approach can be prohibitively expensive. Moreover, because of a lack of strong spectral features and increase in noise due to thermal radiation coming from the sky, it is difficult for spectroscopy to identify galaxies in the redshift range $1.5 \leq z \leq 2.5$, which has been termed the ``redshift desert". 

An alternate method of determining redshifts is to use broadband photometric information to locate broad features in galaxy spectra. The idea originates with Baum (1962), who used photometry in nine bands to locate the 4000\AA\ break. Others (Koo 1985, Loh \& Spillar 1986) generalized the technique and it has become popular recently as a method to estimate the redshifts to galaxies using either the 4000\AA\ break or the Lyman break (see, for example, Connolly \etal 1995, Gwyn \& Hartwick 1996, Sawicki, Lin, \& Yee 1997, Giavalisco 2002 for a review). Photometric redshifts are less precise than spectroscopic redshifts, but have been shown to be reasonably accurate, with a $|z_{spec} - z_{phot}|/(1+z)$ typically much less than 0.1 (Hogg \etal 1998, Wuyts \etal 2009). Although less accurate and prone to catastrophic errors, photometric redshifts have the advantage of being done much more quickly and for fainter galaxies than their spectroscopic counterparts, and can be esily be determined in the redshift desert.

\begin{figure}
\includegraphics[width=8cm]{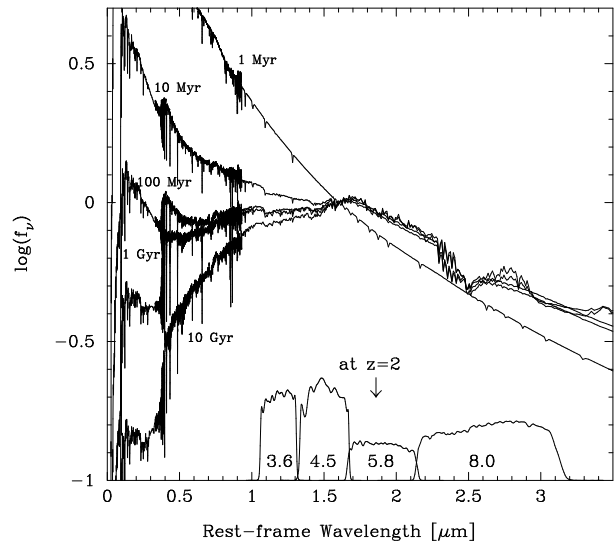}
\caption{\label{fig:bump}Model spectral energy distributions from Bruzual \& Charlot (2003). Shown are SEDs of solar metallicity stellar populations that are forming stars at a constant rate with a Salpeter initial mass function. The SEDs are normalized at 1.6\um. The \bump\ is a prominent feature in all but the very youngest stellar populations. Shown at the bottom of the plot are the filter transmission curves of the IRAC filters redshifted to show them at $z=2$ in the rest-frame of the SEDs.}
\end{figure}

While photometric redshifts have traditionally been done with features detectable at optical or near infrared (NIR) wavelengths (the Lyman break and the 4000\AA\ break), recent deep surveys with information at infrared (IR) wavelengths, such as the Great Observatories Origins Deep Survey (GOODS, Dickinson \etal 2001, Giavalisco \etal 2004), have made other features accessible for use, specifically, the spectral ``bump" at 1.6\um\ (see Figure \ref{fig:bump}).  A nearly ubiquitous feature in all stellar populations, this bump is caused by a minimum in the opacity of the H$^-$ ion present in the atmospheres of cool stars and can be expected to provide a means of estimating redshifts to galaxies (Wright \etal 1994, Simpson \& Eisenhardt 1999, Sawicki 2002, Papovich 2008).

\subsection{Motivation}

The near universality of the \bump\ should make it possible to use it to select highly complete and unbiased samples of galaxies. This is of great significance, as current selection techniques, such as Lyman Break selection (Steidel \etal 1999), Distant Red Galaxy selection (Franx \etal 2003), and $BzK$ selection (Daddi \etal 2004), all require photometry in rest-frame ultraviolet (UV) or optical. The UV/optical flux of a galaxy, however, comes predominantly from hot, young stars, which have relatively short lifetimes. Hence, these current techniques could introduce a bias by favoring galaxies that currently have ongoing star formation and missing a population of passive galaxies. 

The observation of the \bump\ in the infrared makes it well suited to study galaxies at redshifts greater than 1.5, precisely the regime where spectroscopy becomes increasingly difficult. This is also the epoch in which star formation in massive galaxies begins to shut down in the SFR density plot, and is therefore crucial to our understanding of the stellar mass formation history of the universe. The ability to select an unbiased and highly complete catalog of galaxies at this epoch is needed in order to place the best constrains on the stellar mass assembly of the universe. In this, the \bump\ should be a valuable tool.

Furthermore, the James Webb Space Telescope (JWST), now under construction, will provide data in the 0.6-27\um\ range. JWST's lack of sensitivity at shorter wavelenghts means that many of the currently popular selection techniques, while adaptable to higher redshifts, will not be usable with JWST for galaxies around the $z$$\sim$2 peak of the cosmic star formation history. However, the \bump\ is well suited to utilize JWST to study galaxies over a wide range of redshifts, including $z$$\sim$2. It is therefore important to develop the technique now, in order to have an understanding of both its advantages and limitations. Currently, this can be done using the Infrared Array Camera (IRAC, Fazio \etal 2004) instrument aboard the Spitzer Space telescope, which observes in the 3-9\um\ range and thus brackets the \bump\ for $1.3 \lesssim z \lesssim 3$, exactly the period that is of interest.  

The aim of this work is to test the feasibility and limitations of using the \bump\ as a photometric redshift indicator and selection technique and to make an independent, unbiased measurement of the stellar mass function and stellar mass density at $z=2$. In all things, we tried to use only photometric information from bandpasses near the \bump\, and to achieve as much as possible with as little as possible. This paper is divided as follows. Section~\ref{photometry} describes our method of obtaining infrared photometry. Section~\ref{photoz} describes how we determined the photometric redshifts to galaxies using the \bump\ and compares our results with spectroscopy to ascertain an estimate of the quality of the photometric redshifts. In \S~\ref{SMF}, we construct stellar mass functions and compare our results with those from other techniques. Section~\ref{conclusions} lists our conclusions, and provides advice for those who would try to use the \bump\ technique in the future. In all calculations, we use the AB flux normalization (Oke 1974) and assume a cosmology of $\Omega_M = 0.3, \Omega_\Lambda = 0.7,$ and $H_0 = 70$ km/s/Mpc.

\section[Photometry]{Photometry}
\label{photometry}

\subsection{Data}
\label{photometry:data}

We use publicly available data from the Great Observatories Origins Deep Survey (GOODS, Dickinson 2001, Giavalisco 2004), which covers approximately 320 square arcminutes in two fields (North and South).  The Spitzer Space Telescope Legacy Program has carried out deep infrared (IR) observations in these fields with the Infrared Array Camera (IRAC, Fazio \etal 2004)  in four bandpasses (3.6, 4.5, 5.8 and 8.0\um).  All four channels were observed simultaneously, with channels 1 and 3 (3.6 and 5.8\um) covering one pointing on the sky and channels 2 and 4 (4.5 and 8.0\um) covering another.  This 2x2 mapping pattern leads to a small overlap area in each of the North and South fields  between the two filter pointings of about 3 arcminutes. 

In order to cover the whole GOODS region in all four bands, observations were made in two epochs such that the area covered by channels 1 and 3 in the first epoch would then be covered by channels 2 and 4 in the second epoch. The mean total exposure time of the observations is 23 hours in each bandpass, except in the overlap region where exposure time is effectively doubled. 

The unsurpassed depth of these fields at these wavelengths make them well suited to our purposes. However, the relatively large point spread functions (PSF) of the images, with a full-width at half-maximum (FWHM) of approximately 2 arcseconds, prove challenging for extracting photometry. The large PSF arises due to the small mirror size of Spitzer (0.85 m) combined with the large diffraction of light at infrared wavelengths. The crowding in the images is a significant problem, as many galaxies are contaminated by flux from neighbouring objects.  Great care must be taken to properly account for this contamination.

\subsection{Photometry Estimation in Crowded Fields}
\label{photometry:crowd}

The details of our photometric procedure are given in the Appendix and here we give a brief summary of the pertinent points. 
Following the work of others (for example, Fernandez-Soto \etal 1999, Labb{\'e} \etal 2006, de Santis \etal 2007, Laidler \etal 2007), we use high resolution, shorter wavelength data to guide the separation of blended fluxes in low resolution IRAC images, but with a few modifications that we found gave slight improvements.  Essentially, the photometric procedure assumes that galaxies that are confused in the low resolution, longer wavelength image (hereafter the measure image) are resolved in a higher resolution, shorter wavelength image (hereafter the detection image). 

\begin{figure}
\includegraphics[width=8cm]{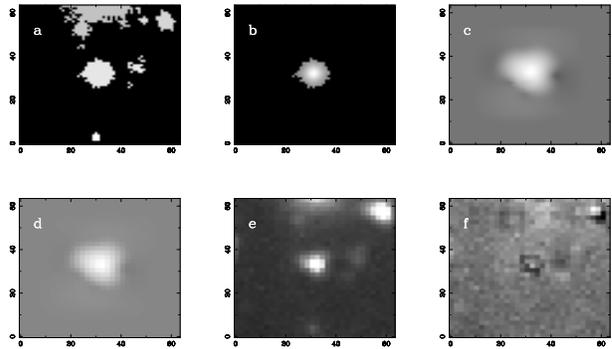}
\caption{\label{fig:msteps}An example of the kernel-fitting photometry procedure. Panel a) shows the SExtractor segmentation map of the detection image, which defines which pixels belong to which galaxy and is used to extract the galaxy of interest and mask other galaxies (shown in Panel b). The detection galaxy is background subtracted and then convolved with the transformation kernel (Panel c) and the pixels are rebinned (Panel d) to match that of the measure image. Panel e) shows the galaxy in the measure image after its background has been subtracted. Finally, Panel f) shows the residual flux remaining after the scaled model has been subtracted from the measure image. This demonstrates that the model is a good match to reality as the residual is on the order of the noise. }
\end{figure}

The process of using the detection image to constrain photometry in the measure image is as follows: (1) Objects are first defined in the IRAC-2 (4.5\um) image using SExtractor. (2) Counterparts in the shorter-wavelenght detection image are then identified, or --- if they are too faint to be seen --- artificially generated. (3) Each galaxy in the detection image is convolved with a transformation kernel in order to match the PSF of the measure image. (4) The convolved galaxies are normalized to unit flux, yielding a model profile for each galaxy in the measure image. (5) The normalized model profiles are each scaled simultaneously to obtain a best-fit to the measure image. An illustration of this process is shown in Fig.~\ref{fig:msteps}.

In our case, the high resolution images consisted of publicly available VLT/ISAAC $K_s$-band data\footnote{The ISAAC observations have been carried out using the Very Large Telescope at the ESO Paranal Observatory under Program ID(s): LP168.A-0485.} 
in the South (Retzlaff \etal 2006) and publicly available Subaru/Suprime-cam $z$-band data (central wavelength 0.85\um) in the North field (Capak \etal 2004). Although it would have been preferable to have $K$-band data in both fields in order to keep all our observations at NIR wavelengths and also to minimize morphological differences between the high-resolution and low-resolution images (
see Appendix), but no near-IR images of GOOD-South were available in the public domain.  It is important to stress that our photometric catalog is based on SExtractor selection in the IRAC2 band (4.5\um), and the shorter-wavelength "detection images" are used only as the basis of the subsequent kernel-convolved photometry at other wavelengths.

In the overlap regions, where data were taken at two different epochs, the rotation of 180$^\circ$ between epochs causes the (asymmetric) Spitzer PSF to be oriented differently in each set of images. Rather than trying to combine the images from the two epochs, we chose to work with each epoch separately, averaging the resulting photometry in the final catalog. Having two independent measurements of each galaxy in the overlap region provides us with a realistic estimate of our random photometric uncertainties. Using the standard deviation of the difference of these two measurements, we find typical uncertainties of 0.09, 0.09, 0.20, and 0.21 mag for the 3.6, 4.5, 5.8 and 8.0\um\ bands respectively for galaxies in our final catalog. (See Appendix.)



\subsection{The Photometric Catalog}
\label{photometry:catalog}

\begin{figure}
\includegraphics[width=8cm]{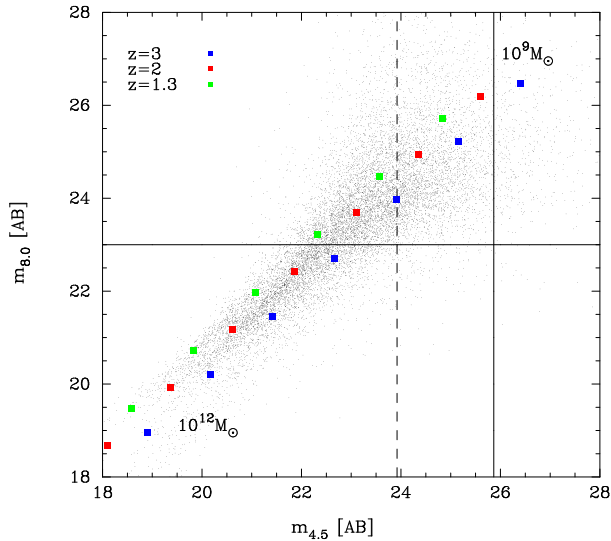}
\caption{\label{fig:limit}Black points represent galaxies with photometry in all four bands. The solid horizontal line shows our empirical hard-limit in the IRAC-4 band for galaxies included in the catalog ($m_{8.0}<23$). Vertical dashed and solid lines show 10$\sigma$ and 3$\sigma$ limiting magnitudes at 4.5\um. Colored squares show model magnitudes for a 1 Gyr solar metallicity galaxy with a constant star formation rate and various stellar masses ranging from $10^9\Msun$ to $10^{12}\Msun$ in decade steps of 0.5. Our IRAC-2, IRAC-4 selected catalog should therefore be complete to approximately $10^{10.25}\Msun$.}
\end{figure}

The result of our photometry procedure was an IRAC-2 selected catalog with approximately 35,000 objects. However, in order to achieve the best results with our limited number of bandpasses, we required that an object be detected in all four IRAC bandpasses. In addition, because of the large photometric scatter in the 8.0\um\ bandpass and the importance of this bandpass as an ``anchor" in determining the photometric redshift (see \S \S\ \ref{photoz:premise}-\ref{photoz:results}), we further restricted the catalog to objects with an 8.0\um\ magnitude less than 23. Our catalog is thus a joint IRAC-2, IRAC-4 selected catalog. Because the \bump\ is caused by older stars (which is where most of the stellar mass in a galaxy lies), it could also be said that this is very nearly a stellar mass selected catalog (at least for redshifts greater than 1.5, which is where our interest lies). Figure \ref{fig:limit} shows the IRAC-2, IRAC-4 color space, our 10$\sigma$ and  3$\sigma$ IRAC-2 limiting magnitudes, as well as model magnitudes at various redshifts showing what stellar masses of galaxies we should expect to be included in our catalog. We find that our catalog to be complete to approximately $10^{10.25}\Msun$.

Next, we used the Grazian \etal (2006) and Barger \etal (2009) catalogs to remove objects catalogued as stellar.  Finally, we consider the issue of active galactic nuclei (AGN), which could pose a problem to our SED fitting procedure (\S\ \ref{photoz:SED}) as their colors will be vastly different from the model SED's. Although there has been work done in trying to select AGN using only the IRAC bands (Lacy \etal 2004, Stern \etal 2005, Alonso-Herrero \etal 2006) these techniques also select a high number of galaxies without AGN. Although we could try to use observations at x-ray or radio wavelengths to detect AGN, that would be contrary to the spirit of this work (mainly, what can be accomplished using only bands around the \bump). In the end, we decided not to attempt to filter out AGN contamination. This should not be of great concern, as Barger \etal (2009) point out that AGN contamination is a small effect, on the order of a few percent. 

In summary, the final sample adopted in this work consists of 5557 objects with photometry in all four IRAC bands and an 8.0\um\ magnitude less than 23. This catalog covers an area of 303.8 arcmin$^2$ and is approximately equivalent to a stellar mass-selected sample that reaches to $10^{10.25}\Msun$ at $z=2$.


\section[Redshift Selection]{Redshift Selection}
\label{photoz}

In this section we discuss how the \bump\ can be used to select galaxies at specific redshifts given photometry in the four IRAC bands. In \S\ \ref{photoz:premise} we describe the basic premise of why the \bump\ is an indicator of redshift. In \S\ \ref{photoz:SED} we describe our method of fitting model spectra to the data. In \S\ \ref{photoz:models} we describe the models in more detail, as well as discussing some of the degeneracies and limitations of trying to model the \bump. In \S\ \ref{photoz:results} we show the results of fitting the models to the data that have confirmed spectroscopic redshifts and discusses the degrees of contamination and completeness. 

\subsection{Premise}
\label{photoz:premise}

\begin{figure}
\includegraphics[width=8cm]{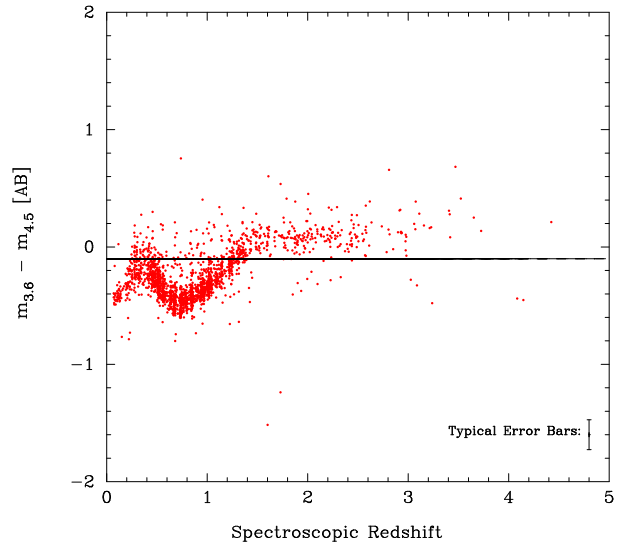}
\caption{\label{fig:zcplot_12}Color between the first two IRAC bands as a function of spectroscopic redshift for galaxies in our photometric catalog. Colors are typically blue at lower redshifts and turn red at higher ones so that nearly all galaxies with $z>1.3$ have a color greater than -0.1.}
\end{figure}

As the \bump\ changes wavelength with redshift and passes through two adjacent bandpasses, the color between those bandpasses will change from blue to red. For example, at redshift 1.3, the \bump\ has been shifted to a wavelength of $\sim$3.7\um\ and is just entering the region between the 3.6 and 4.5\um\ bandpasses. The change in color with redshift can be seen in Figure \ref{fig:zcplot_12}, where the color between the 3.6 and 4.5\um\ bandpasses is plotted against spectroscopic redshift for galaxies in both GOODS fields. The colors of galaxies at redshift less than 1.3 are typically blue, but then become red and remain that way out to higher redshifts. By redshift $\sim$1.5, nearly all galaxies have a [3.6]-[4.5] color greater than -0.1. Consequently, this color can be used to select a largely complete catalog of high redshift galaxies, although with a fair amount of contamination from low redshift galaxies (Papovich 2008).

\begin{figure}
\includegraphics[width=8cm]{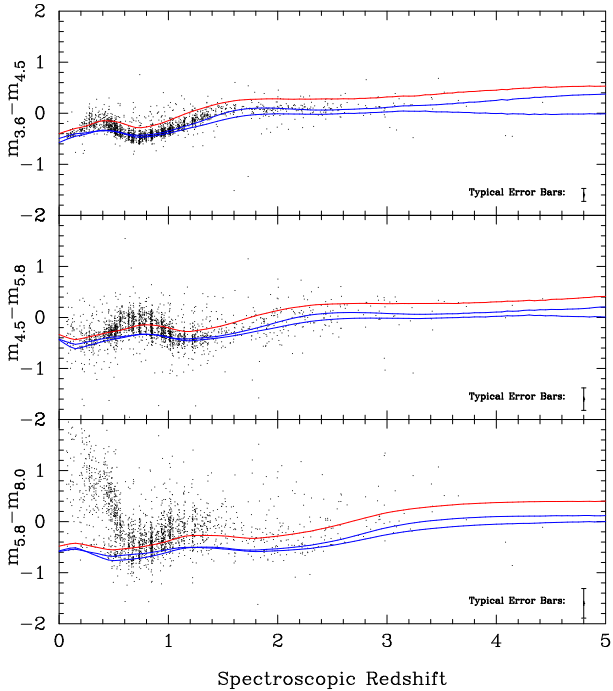}
\caption{\label{fig:zcplot}Colors between adjacent IRAC bands as a function of spectroscopic redshift for galaxies in our photometric catalog. The solid blue lines represent the color range of the models we use to fit a photometric redshift. The solid red curve shows how discrepancies at low redshift could be explained by extremely dusty starbursts or LIRGS (see \S\ \ref{photoz:models}).}
\end{figure}

Figure \ref{fig:zcplot} shows how the same effect can be seen at longer wavelengths as the \bump\ continues to shift in wavelength. Here, a similar pattern can be seen in all three colors, but with the features shifted to higher redshifts at longer wavelengths. The change in color from blue to red happens at $z\approx1.3, z\approx2$, and $z\approx3$ for the [3.6]-[4.5], [4.5]-[5.8], and [5.8]-[8.0] colors respectively, although photometric scatter and a lack of high redshift objects with spectroscopy make it hard to distinguish at the longest wavelengths. The IRAC bands should therefore be able to effectively determine the redshift to galaxies in the range $1.3 \leq z \leq 3$. At redshifts less than this, a strong wiggle caused by the CO absorption band causes redshifts to be degenerate, and at higher redshifts the \bump\ has been shifted beyond the last bandpass and so no redshift information can be determined. The blue curves show the range of colors of our model templates and will be discussed in more detail in \S\ \ref{photoz:models}. 

\begin{figure}
\includegraphics[width=8cm]{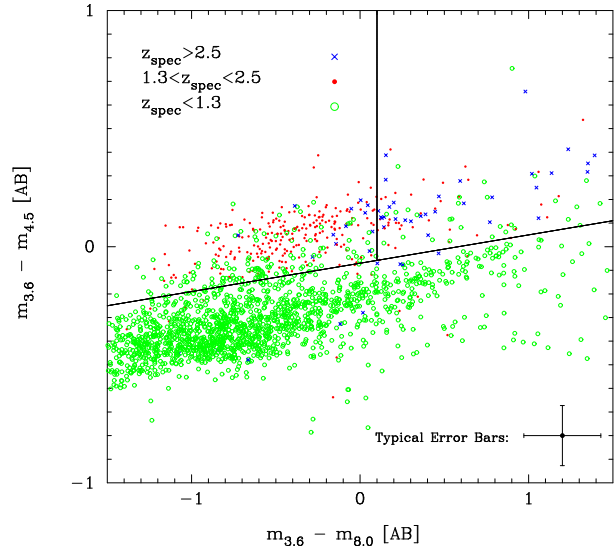}
\caption{\label{fig:ccplot}Color-color diagram showing how IRAC colors can be used to select high redshift galaxies. Green dots have spectroscopic redshifts less than 1.3, red 1.3-2.5 and blue greater than 2.5.}
\end{figure}

In principle, one could use simple color criteria to select galaxies in a specific redshift range. For example, as mentioned above, the [3.6]-[4.5] color is excellent at seperating galaxies at redshifts greater than 1.3. Similarly, the \bump\ will lie between 3.6 and 8.0\um\ filters at a redshift of about 2.5, and so this color could be used as an upper limit. Shown in Figure \ref{fig:ccplot} is the color-color diagram for galaxies in our photometric catalog with spectroscopic redshifts. Dividing lines have been drawn at $m_{3.6}-m_{4.5} = 0.12(m_{3.6}-m_{8.0} )-0.07$ and $m_{3.6}-m_{8.0}=0.1$ to make three regions: a $z<1.3$ region at the bottom, $1.3 \leq z \leq 2.5$ in the upper left, and a higher redshift region in the upper right. Galaxies with spectroscopic redshifts in the range of interest (red points) typically fall in the upper-left region with a high degree of completeness (221/291 or $\sim$75\%), although there is a substantial amount of loss (41/291 or $\sim$14\%) due to scatter into the "higher redshift"  area. This scatter could simply be due to photometric scatter (particularly in the 8.0\um\ band), but could also be caused by large amounts of dust (as discussed in \S\ \ref{photoz:models}) or the presence of AGN. There are also a significant number of low redshift galaxies contaminating the upper left region (127 out of the 353 galaxies in that region or $\sim$36\%). Papovich (2008) suggests that these contaminates are a population of infrared luminous star-forming galaxies in the range $0.2 < z < 0.5$. Under this scenario, the red [3.6]-[4.5] color is a result of warm dust heated by star formation (see also Imanishi 2006). 

The degree of contamination from low-$z$ galaxies can be lowered by increasing the intercept of the $m_{3.6}-m_{4.5}$ criterion, but at the cost of sacrificing completeness. For example, if the color criterion is changed to $m_{3.6}-m_{4.5} = 0.12(m_{3.6}-m_{8.0})+0.02$, the contamination level from low redshift galaxies drops to $\sim16\%$, while the completeness level drops to $\sim65\%$.  

For color-color selection, the \bump\ works much better as a lower limit than as an upper limit. Of all galaxies with redshifts greater than 1.3, 310/343 or $\sim$90\% lie above the line $m_{3.6}-m_{4.5} = 0.12(m_{3.6}-m_{8.0}) -0.07$. The contamination from low redshift galaxies in this entire upper region is 158/468 or just over 33\%. Again, the degree of contamination can be reduced by adjusting the intercept of the $m_{3.6}-m_{4.5}$ criterion. Similar results with a slightly different IRAC color selection technique were obtained by Barger \etal (2009) in the North field alone, although with a slightly higher contamination rate ($\sim$40\%). It is worth mentioning that these estimates of contamination fraction are upper limits, as spectroscopic catalogs are most likely incomplete at higher redshifts.

The selection of galaxies using these three bands (hereafter IRAC selection) is readily comparable to the popular $BzK$ selection technique (Daddi \etal 2004). Both techniques use three bands and are able to select both star-forming and passive galaxies at high redshift ($1.4 < z < 2.5$ for $BzK$, $z > 1.3$ for IRAC color selection), although IRAC selection cannot distinguish between star-forming and passive galaxies. While some groups have tried to test the reliability of $BzK$ selection using large samples of photometric redshifts (\eg Kong \etal 2006, Grazian \etal 2007, Quadri \etal 2007), the difficulty with this approach is that the colors plotted in the $BzK$ diagram are the very same used to determine the photometric redshifts. Independent spectroscopic redshifts provide the best validation of any color selection technique, and recent work has been done to estimate the completeness and contamination of $BzK$ selection using spectroscopic catalogs (Barger \etal 2008, hereafter B08, Popesso \etal 2009, hereafter P09). 

The completeness level in the star forming region of the $BzK$ diagram is found to be 88\% in B08 and 86\% in P09. This is slightly less than the 90\% completeness level for IRAC selection found in this work and B08. While Daddi \etal (2004) originally stated the contamination level of $BzK$ at 12\%, P09 found the contamination in the star forming region to be 33\% (23\% from $z < 1.4$ galaxies and 10\% from $z > 2.5$ galaxies), and B08 found it to be a minimum of 33\% and a maximum of 64\% (all from $z < 1.4$ galaxies). IRAC selection therefore seems to perform as well as $BzK$ selection's star-forming criterion.

Completeness and contamination in the passive $BzK$ region are not well constrained due to small number statistics. Most passive galaxies are very faint in the $B$ band, often below the limiting magnitude of large surveys. These galaxies should be much more prominent in the IRAC bandpasses. It should be mentioned as well, that B08 found that not all $BzK$ selected galaxies with spectroscopic redshifts $z_{spec} > 1.4$ were also selected with IRAC colors or vice-versa, and that the most complete catalog was comprised of galaxies that satisfied either one or the other selection criteria. 
 
 Although color selection can be very useful for selecting galaxies in a certain redshift range, the amount of information that can be extracted from two bands at a time is limited. In this work, we take into account information from all bands simultaneously by fitting model spectra to the photometry. This process is described in the next section.
 
\subsection{SED fitting}
\label{photoz:SED}

Photometric redshifts are estimated by comparing observed broadband photometry with grids of model templates. The models can vary in redshift, star formation history, amount of extinction, metallicity, age of the stellar population, stellar initial mass function (IMF), \etc\ In this work, we limited our parameters to stellar age and redshift as we found that there was not enough information contained in the IRAC bands to constrain the other parameters. However, as discussed further in section \S\ \ref{photoz:models}, the shape of the \bump\ does not yield much information on the age of a galaxy, and thus the age span of our model templates was restricted in order to minimize effects of incorrect age estimates.

We used as our basis model spectra from the 2003 version of the GISSEL spectral synthesis package (Bruzual \& Charlot 1993, 2003), with a single stellar population, Salpeter (1955) IMF, 0.2 solar metallicity, zero extinction, and ages ranging from 0.3-3 Gyr (see \S\ \ref{photoz:models} for an explanation as to why we feel these are reasonable choices). 

Using the SEDfit software package (Sawicki \& Yee 1998, Sawicki 2010 [in prep]), these model spectra were redshifted onto a grid of redshifts spanning $0 \leq z \leq 5$ in steps of 0.05 and attenuated using the Madau (1995) prescription for continuum and line blanketing due to intergalactic hydrogen along the line of sight. Finally we integrated the resultant observer-frame model spectra through filter transmission curves to produce model template broadband fluxes. In order to match the model fluxes to observations, for each object the software compared the observed fluxes with each template in the grid by computing the statistic
\begin{equation}
\chi^2 = \sum_{i} {(f_{obs}(i) - sf_{tpt}(i))^2\over\sigma^2(i)},
\end{equation}
where $f_{obs}(i)$ and  $\sigma$(i) are the observed flux and its uncertainty in the $i$th filter, and $f_{tpt}(i)$ is the flux of the template in that filter. The variable $s$ is the scaling between the observed and template fluxes, and can be computed analytically by minimizing the $\chi^2$ statistic with respect to $s$ giving
\begin{equation}
s = {\sum_{i} {f_{obs}(i)f_{tpt}(i) / \sigma^2(i)}\over{\sum_{i}{f_{tpt}^2(i) / \sigma^2(i)}}}
\end{equation}
(Sawicki 2002). For each object, the most likely redshift is determined by the smallest $\chi^2$ value over all the templates.

Photometric redshifts are prone to catastrophic errors because of degeneracies in the model templates, and also because of spectral slope information lost by integrating over the broadband filter range. In the next section, we attempt to analyze and understand possible causes for catastrophic failures. 

\subsection{Models}
\label{photoz:models}

\begin{figure}
\begin{center}
\includegraphics[width=8cm]{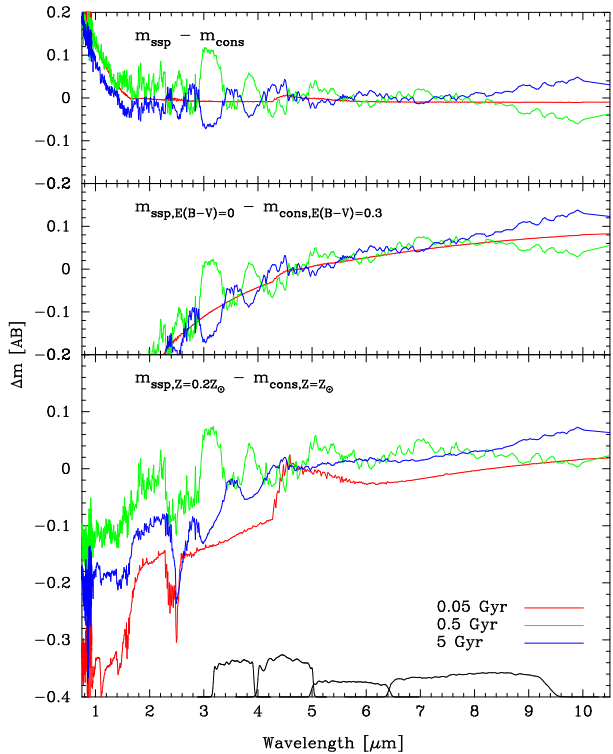}
\caption{\label{fig:robust}Magnitude difference of various models as a function of wavelength. The plots show the effect of assuming different star formation histories (either a single stellar population (SSP) or constant star formation (cons)), extinction (E(B-V) of either 0 or 0.3), and metallicity (\Zsun or 0.2\Zsun) at ages 0.05, 0.5, and 5 Gyr. All models are from Bruzual \& Charlot (2003) and have been redshifted to $z=2$, reddened using the Calzetti (2000) extinction law and attenuated using the Madau (1995) prescription. The models have been normalized to flux at the location of the \bump\ at this redshift (4.8\um). Shown at the bottom in black are the locations of the four IRAC filters. Note that in all cases, the choice of model has an effect typically less than 0.1 mag over the IRAC bandpasses, but which increases drastically at shorter wavelengths.}
\end{center}
\end{figure}

The shape of the \bump\ is very robust, in that it does not depend greatly on the galaxy's star formation history, dust content, or metallicity. This is demonstrated in Figure \ref{fig:robust}, which shows the differences in model spectra with varying parameters for a galaxy at redshift 2. The spectra have been normalized by the flux at the wavelength of the \bump\  at that redshift. Note that the differences are typically less than 0.1 mag for IRAC wavelengths, but then diverge rapidly for bluer wavelengths. Because of this robustness, our choice of parameters for the model templates should not have a great effect on the best fit redshift, at least for galaxies near redshift 2, thus reducing the possibility of systematic errors due to poor choice of input parameters. The robustness also removes many of the degeneracies inherent in SED fitting. For example, at blue wavelengths, the extinction due to dust causes a galaxy's spectrum to appear redder. However, a similar redder appearance can be produced by an old galaxy with a star formation history close to that of a single stellar population. It can be difficult to distinguish between these two effects and so a galaxy could be assigned a high extinction value when it merely has an older stellar population or vice-versa. This degeneracy does not exist with the \bump, whose shape does not change dramatically with these parameters. The robust shape means, however, that no information can be determined about these parameters from the model fits. In general though, we consider the limited model parameter space of this technique to be a benefit.

It is unlikely that the choice of IMF will drastically alter redshift results, as changing the amount of massive stars relative to cooler stars will have similar effects as a change in star formation history. Nor does the age of the stellar population have a great effect on the shape of the \bump\ (see Figure \ref{fig:bump}). Only in extremely young populations is it obscured by the power law from the youngest, brightest, most massive stars. This power law makes the redshifts of extremely young galaxies degenerate. As Sawicki (2002) points out, care should be taken with galaxies that have a best-fit age of less than $\sim$0.01 Gyr. This is a very small percentage of our catalog (less than 0.1\% when fitting all model ages).

While parameters discussed above do not drastically alter the shape of the \bump, and hence will not affect the best-fit redshift of galaxies around redshift 2, it is important to note that they may have an effect on the estimated stellar mass of that galaxy. The systematic bias of the estimated galaxy masses introduced by our choice of extinction or star formation history is most likely not significant since the large majority of the stellar mass contained within a galaxy is due to older, cooler stars. These are exactly the stars that are probed by the \bump\ and thus a mismatch at bluer wavelengths should not alter the predicted mass greatly. 

\begin{figure}
\includegraphics[width=8cm]{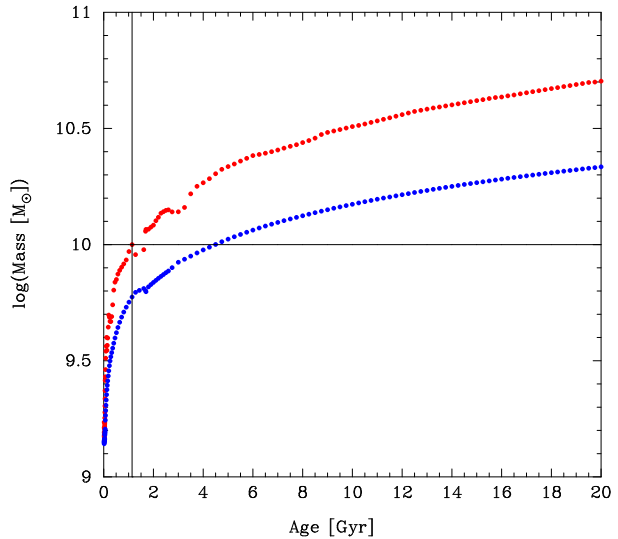}
\caption{\label{fig:agemass} Stellar mass estimation of a fit to an input galaxy with age 1 Gyr, a mass of 10$^{10}\Msun$ and a single stellar population versus fitted model age. The upper (red) points are for model templates with a single stellar population and the lower (blue) points are for constantly star forming model templates. While the error in mass estimation can vary with age by as much as 0.5 dex, the error due to a mismatch in star formation history is typically less than a factor of two at a given age.}
\end{figure}

We found, however, that a mismatch in model ages could produce a measurable systematic error in stellar mass estimation. Because of the similarity in the shape of the \bump\ at nearly all ages and the large photometric uncertainties in our catalog, it is quite likely that many stellar mass estimates could be off by 0.5 dex if all possible model ages are included in the fitting procedure (see Figure \ref{fig:agemass}). To avert this possible systematic bias, we constrained our model ages to range from 0.3 to 3 Gyr. We feel that these are reasonable restrictions, as there should not be many galaxies with a stellar population age less than 300 Myr, and at redshift 2 only approximately 3.5 Gyr had elapsed since the Big Bang. This restriction limits the error in stellar mass estimation to be at most a factor of $\sim2$, which is typical of the accuracy of stellar mass estimates obtained with SED modelling (Kauffmann \etal 2003, Papovich \etal 2006, Fontana \etal 2006, P{\'e}rez-Gonz{\'a}lez \etal 2008). 

Other factors may affect stellar mass estimation, but are not investigated in this work. For instance, proper treatment of stars in the post asymptotic giant branch phase can influence spectral synthesis models and, hence, stellar mass estimates (Maraston \etal 2006, Bruzual 2007). In addition, the choice of IMF could effect things in a systematic way, as mentioned earlier. A detailed investigation of these effects is beyond the scope of this paper.

The robustness of model parameter choice relies on the \bump\ being in the wavelength range of the IRAC filters. At redshifts lower than 1.3 or higher than 3, this will not be the case, and our models could lead to catastrophic errors at these redshifts. Referring back to Figure \ref{fig:zcplot}, the blue curves represent the color space probed by our models. In the range of $1.3 < z < 3$, these models appear to adequately represent our photometric catalog, but there are clear discrepancies at lower redshifts. The most prominent of these is the sharp rise of the [5.8] - [8.0] color at redshifts less than 0.6 which is not predicted by the models. This feature is due to the strong polycyclic aromatic hydrocarbon (PAH) emission feature at 6.2\um, which is not included in the models, but greatly increases the flux observed in the 8.0\um\ band. Another effect not predicted by our models is the redder color of galaxies at redshifts $z<1$. This discrepancy is most likely due to the large population of luminous infrared galaxies (LIRGs) observed with Spitzer at $z\leq1$ (Le Floc'h \etal 2005, P{\'e}rez-Gonz{\'a}lez \etal 2005), in which warm dust causes the spectrum to redden at rest-frame wavelengths greater than 2\um\ (Imanishi 2006). To demonstrate that this is likely the case, the red curve in Figure \ref{fig:zcplot} shows a 100 Myr model with a constant star formation rate and extremely high extinction (E(B-V)=0.7) consistent with the dusty star formation expected in LIRGs at low-$z$. Indeed, this model seems to better fit the redder colors at lower redshift. Although there are many LIRGs at $z\sim2$, our simple models are still able to reproduce the colors at this redshift much better than at lower redshifts. The lack of discrepancy between or models and photometry at $z\sim2$ is most likely due to the average attenuation factor at $z\sim2$ being 8-10 times smaller than those at lower redshifts (Reddy \etal 2006b, Burgarella \etal 2007, Buat \etal 2007, Reddy \etal 2008).  

\begin{figure}
\includegraphics[width=8cm]{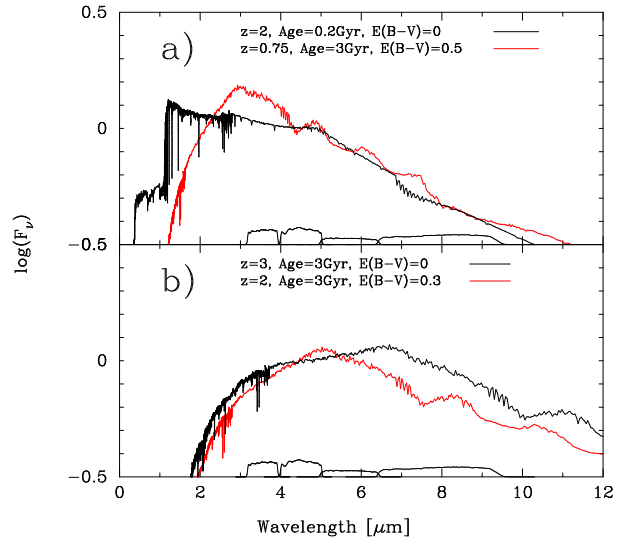}
\caption{\label{fig:spectra}Model spectra (black curves) could be erroneously be fit to older, dusty galaxies at a lower redshift (red curves). The models are as in Figure \ref{fig:robust} and have been scaled to have the same flux at 4.5\um.}
\end{figure}

Improperly modeling low redshift galaxies can lead to a large number of catastrophic redshift errors, with low redshift galaxies often fit erroneously to higher redshifts (see Panel a) of Figure \ref{fig:spectra}). While it could be possible to try and include LIRG SEDs in our model templates, if we are only interested in $z\sim2$ galaxies, we can instead use color criteria to cull the low redshift galaxies. For example, excluding galaxies with a [5.8]-[8.0] color greater than 0.4 effectively removes many of the galaxies at redshifts less than 0.6, and as already mentioned, the [3.6]-[4.5] color efficiently removes galaxies at redshifts less than 1.3. If more accurate photometry were available, it may also be possible to improve model fits to low-$z$ galaxies by constructing empirical model templates from the photometry and spectroscopic redshift information, but this is not attempted in this work.

We also found that it is possible for galaxies in our desired redshift range to be erroneously fit to higher redshifts. This could be the result of our model templates failing to account for all the conditions present in galaxies at this redshift. For example, Daddi \etal (2007) have shown that the presence of hot dust emission in {\it BzK} galaxies at $z=2$ can shift the \bump\ to longer wavelengths. As well, Spitzer/IRS spectroscopic observations show that some $z\sim2$ galaxies have SEDs peaking at 5.8\um\ (Weedman \etal 2006, Farrah \etal 2008, Huang \etal 2009, Desai \etal 2009). This could lead to the presence of a systematic error when deriving photometric redshifts using solely the \bump. 
 
Another likely cause of erroneously high photometric redshifts is simply due to an overestimation of the flux in the 8.0\um\ band. In Panel b) of Figure \ref{fig:spectra}, the importance of this fourth band is demonstrated by showing two different models that would have similar magnitudes in the first three IRAC bands and the only appreciable difference being the 8.0\um\ flux. The importance of this fourth band in determining the redshift cannot be understated:  it can often discriminate between spectral degeneracies in the other three bands.

The addition of lower and higher wavelength bandpasses would, of course, help to better constrain the redshift of the galaxies. Indeed, many surveys have been conducted utilizing a large number (10--14) bandpasses covering the spectrum from $U$ to 24\um\ (Grazian \etal 2006, Reddy \etal 2006, Wuyts \etal 2008). However, this approach requires a great deal of observing time, and it would be preferable to obtain quality results with as little data as possible. Moreover, the addition of extra bandpasses not near the \bump\ would introduce other degeneracies in the models that would have to be taken into account. In this work, we strived to push the limits of what can be accomplished using solely filters around the \bump\ at redshift 2, and we limited ourselves to only the four IRAC filters.

\subsection{Results}
\label{photoz:results}

In this work, our goal was to study galaxies around redshift 2, and to do so, we use a generous photometric redshift range of $1.5 \leq z_{phot} \leq 2.5$, which corresponds to a range in lookback time spanning approximately 1.6 Gyr from $\sim10.9$ Gyr ago to $\sim9.3$ Gyr ago. In this section, we discuss the quality of our photometric redshifts and try to understand the limitations of using the \bump\ to determine them.

Using only the four IRAC bandpasses, we ran our photometric catalog through our SED fitting procedure to obtain photometric redshifts for each galaxy, as well as best-fit ages and stellar masses. We culled from our catalog any galaxies with a fitted redshift greater than 3, as the \bump\ has passed the IRAC bandpasses by this redshift, and color information becomes degenerate (see Figure \ref{fig:zcplot}). Thus, any objects with $z_{phot}>3$ have redshifts which are poorly constrained at best and erroneous at worst. 

\begin{figure}
\includegraphics[width=8cm]{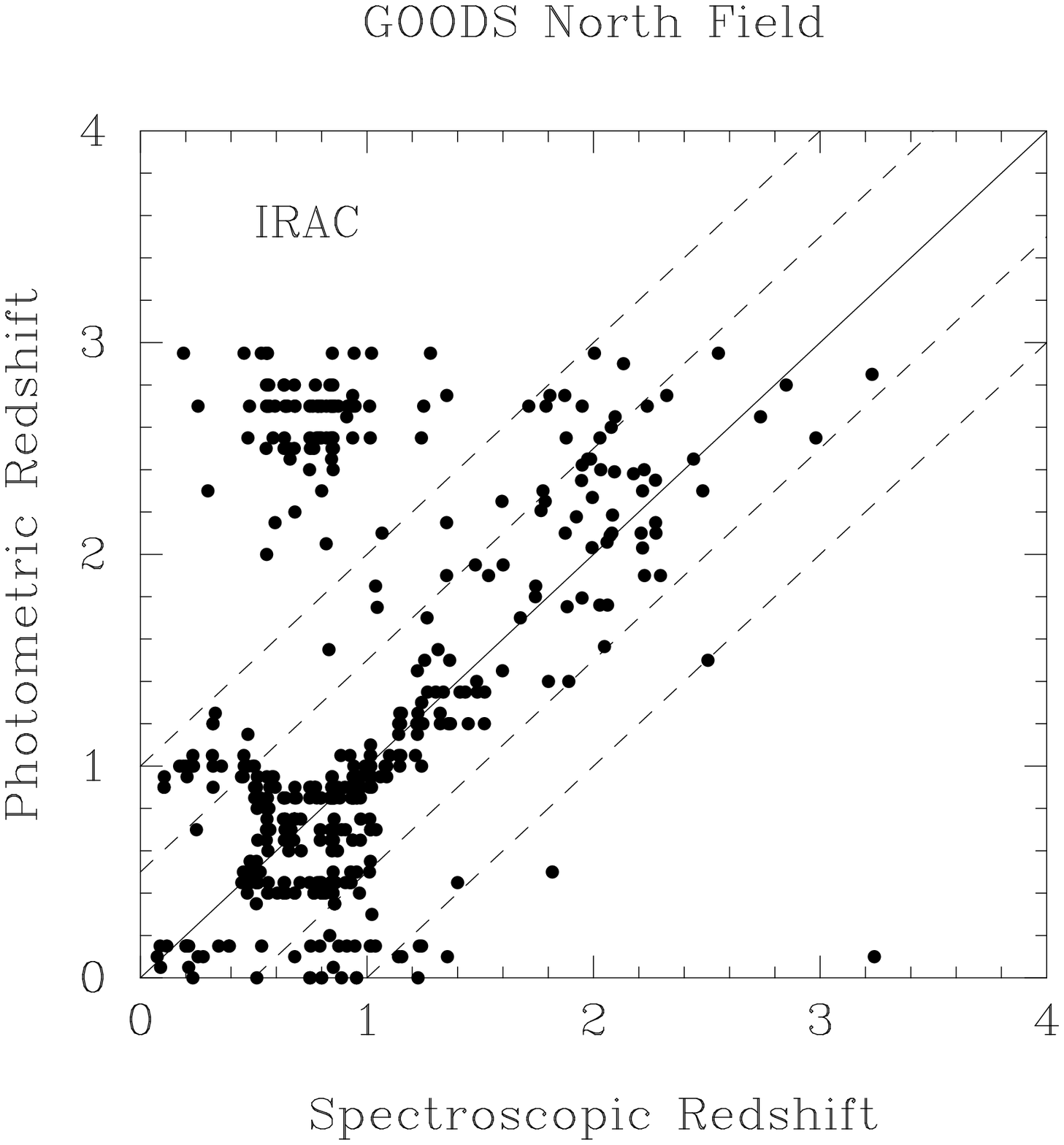}
\includegraphics[width=8cm]{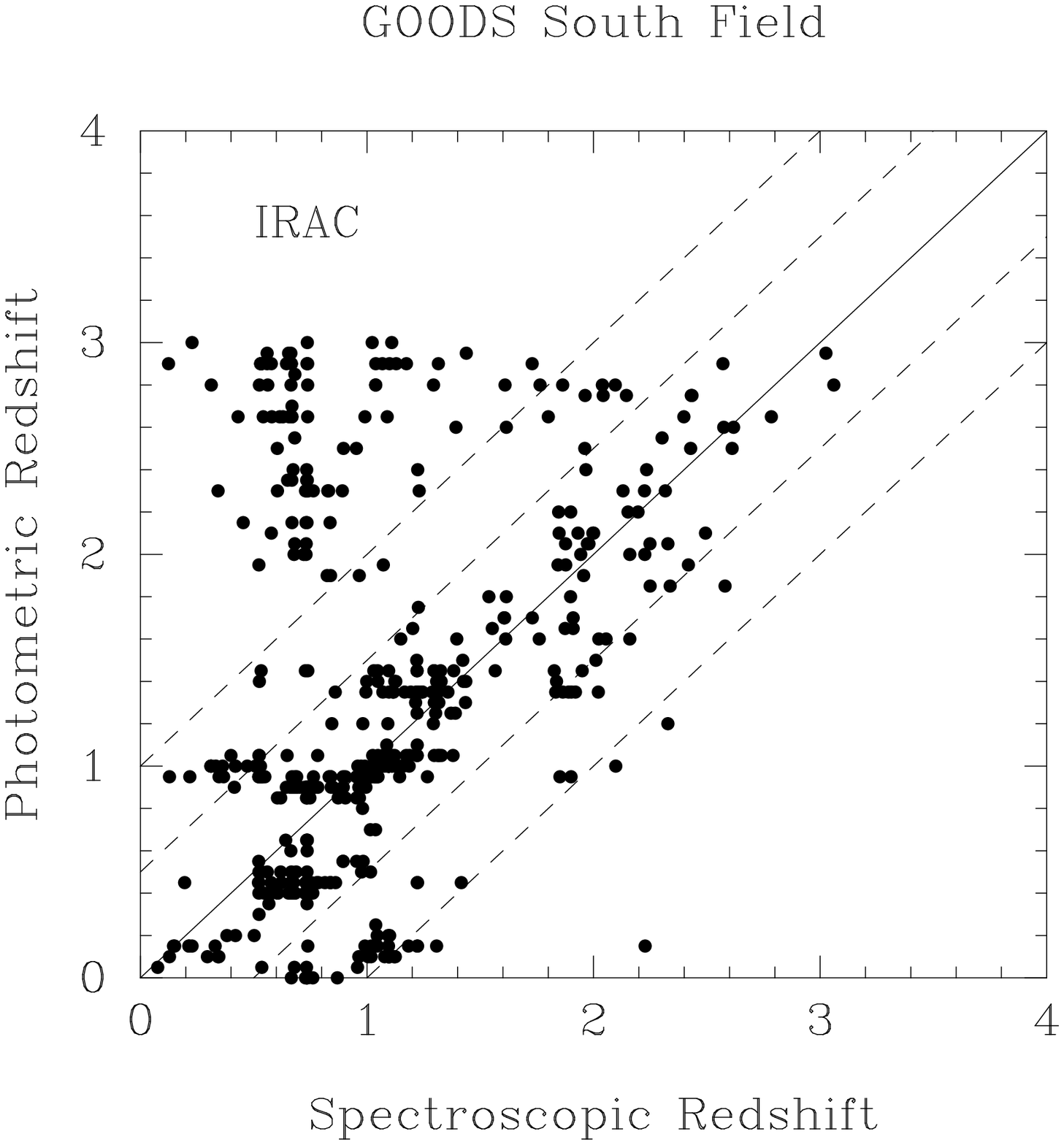}
\caption{\label{fig:zzplot}Photometric redshift as a function of spectroscopic redshift in the North and South fields of the GOODS survey for 462 and 512 galaxies respectively.}
\end{figure}

We compared the remaining galaxies with spectroscopic redshifts where available. The spectroscopic catalog in the South field comes from the GOODS-MUSIC catalog (Grazian \etal 2006), which combines a number of surveys (Wolf \etal 2001, Le F{\`e}vre \etal 2004, Szokoly \etal 2004, Mignoli \etal 2005, Vanzella \etal 2005,  Vanzella \etal 2006), and also recent spectroscopy by Popesso \etal (2009) focusing on galaxies at $1.8 < z < 3.5$. In the North field we use the spectroscopic catalog of Barger \etal (2009), which also made use of several other previous surveys (Barger \etal 2003, Wirth \etal 2004, Cowie \etal 2004, Swinbank \etal 2004, Chapman \etal 2004, Chapman \etal 2005, Treu \etal 2005, Reddy \etal 2006, Trouille \etal 2008,  Barger \etal 2007). Figure \ref{fig:zzplot} shows the result of this comparison. It is apparent there is a great deal of upscatter from lower redshift galaxies for reasons discussed in \S\ \ref{photoz:models}. It should be stressed here, however, that the percentage of outliers in these figures is most likely highly misleading due to probable incompleteness of the spectroscopy at higher redshifts. 

We assume that contamination from high redshift objects being assigned a lower redshift is minimal and negligible compared to the contamination from low redshift objects. The apparent magnitude limits in our photometric catalog likely lead to fewer high redshift objects being included compared to the number of low redshift objects, as only the very brightest high redshift objects will be observable. While it is impossible to confirm this assumption with the low number of spectroscopic observations available at high redshift, we found that, in the small sample of objects (20) in our photometric catalog with $z_{spec}>3$, it was far more likely for redshift to be overestimated than underestimated. None of the 20 high redshift objects were fit to redshifts in the range $1.5 \leq z_{phot} \leq 2.5$.

\begin{figure}
\includegraphics[width=8cm]{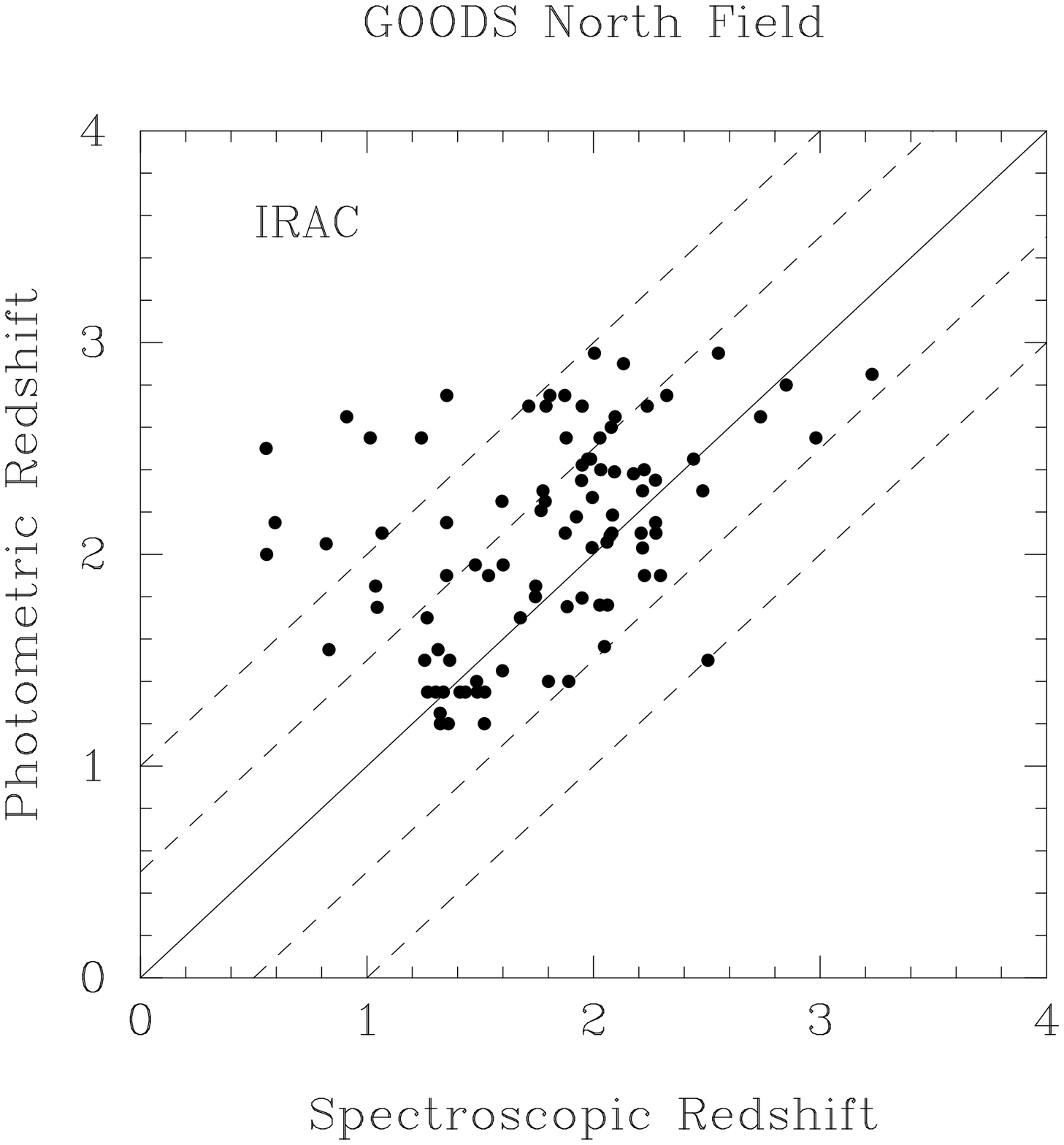}
\includegraphics[width=8cm]{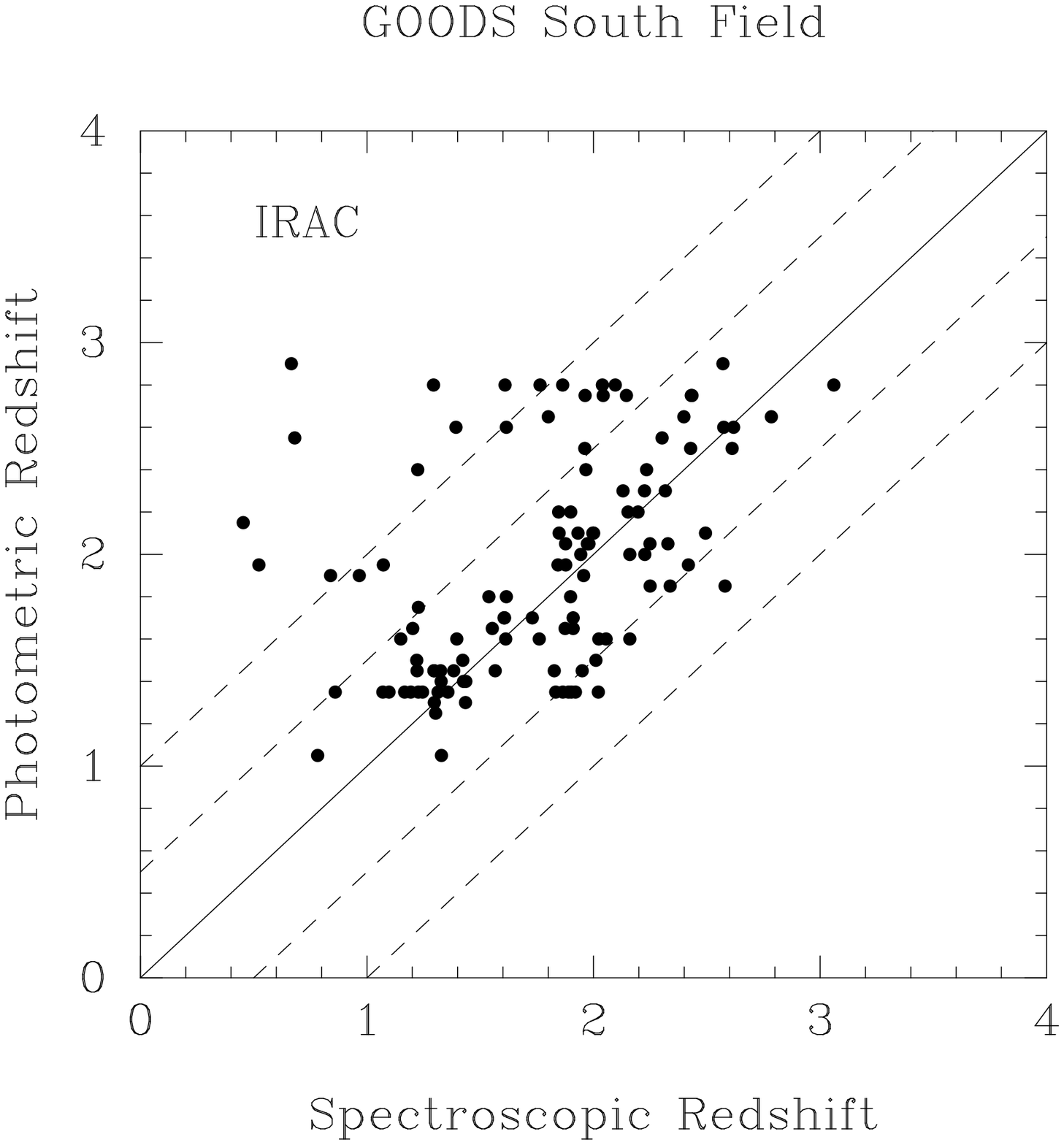}
\caption{\label{fig:cut}Photometric redshift as a function of spectroscopic redshift in the North and South fields of the GOODS survey for 76 and 114 galaxies respectively that remain after a $m_{3.6}-m_{4.5}<-0.1$ cut.}
\end{figure}

We examined various ways of dealing with contamination in our redshift sample. If one is only interested in galaxies at high redshift, then an efficient way of removing low-redshift galaxies without sacrificing completeness is to make an [3.6]-[4.5] color cut of -0.1 (see \S\ \ref{photoz:models}, Papovich 2008). The results of applying this cut are shown in Figure \ref{fig:cut} where $\sim90\%$ (165/182) of the outliers (\ie galaxies that have a $z_{spec}<1.5$ or $z_{spec}>2.5$ but are fitted to a redshift in the range $1.5 < z_{phot} < 2.5$) have been removed while eliminating less than 5\% of the galaxies with spectroscopic redshifts in the range $1.5 < z_{spec} < 2.5$. All of the galaxies with $1.5 < z_{spec} < 2.5$ that were culled with this color cut had best-fit photometric redshifts below 1.5 and so would not have been included in our study group in any case. Other criteria such as best-fit age or $\chi^2$ value were found to also be capable of improving accuracy (\ie by culling any galaxy that has an age less than some particular age, or a $\chi^2$ value greater than some value), but at the cost of significantly sacrificing completeness.

Not pictured in the graphs, a significant fraction of galaxies also get pushed up to redshifts higher than $z_{phot}=3$. Of the 194 galaxies in our photometric catalog with spectroscopic redshifts in the range $1.5 \leq z_{spec} \leq 2.5$, only 86 are in the same region of photometric redshifts, giving a completeness percentage of only $\sim$45\%. The completeness improves, however, for brighter galaxies, increasing to $\sim$70\% (29/42) of galaxies with 4.5\um\ magnitudes less than 21. Of the galaxies with $1.5 \leq z_{spec} \leq 2.5$ that were incorrectly fit to photometric redshifts {\it{outside}} the range $1.5 \leq z_{phot} \leq 2.5$, approximately 20\% were assigned photometric redshifts just below the correct redshift region, while the vast majority ($\sim$80\%) were upscattered to a higher redshift. This is most likely due to photometric scatter in the 8.0\um\ band causing the long wavelength flux to be overestimated. Although AGN or misfits due to model assumptions cannot be ruled out in all cases, it is necessary to stress the importance of having accurate photometry at longer wavelengths in determining photometric redshifts using the \bump. 

For the galaxies that were fit to a model of $z_{phot}>3$, we redid the fitting procedure but omitted the 8.0\um\ band. We found that doing so was able to increase the completeness of galaxies in the $1.5 \leq z_{spec} \leq 2$ region, confirming the conjecture that the 8.0\um\ band pass was to blame for the upscatter. Fits to galaxies with larger spectroscopic redshifts were not improved because colors in the first three IRAC bands become degenerate. This iteration of the fitting procedure was especially effective in the North field as opposed to the South, which could hint at possible problems in using the $z$-band as a detection image for the 8.0\um\ measure image. The issue may be caused by morphological differences in galaxies between these two well separated wavelengths.

\begin{figure}
\includegraphics[width=8cm]{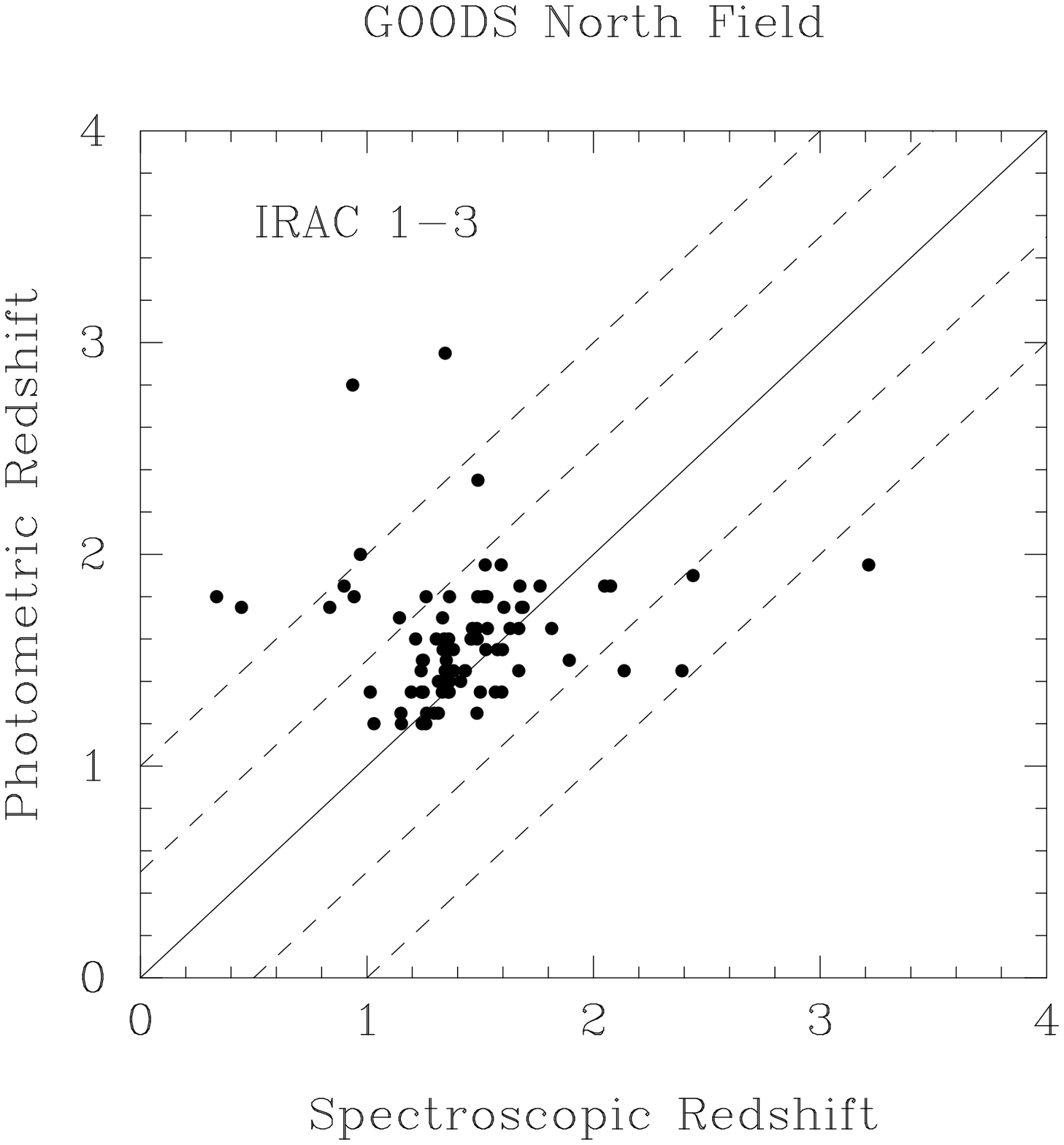}
\includegraphics[width=8cm]{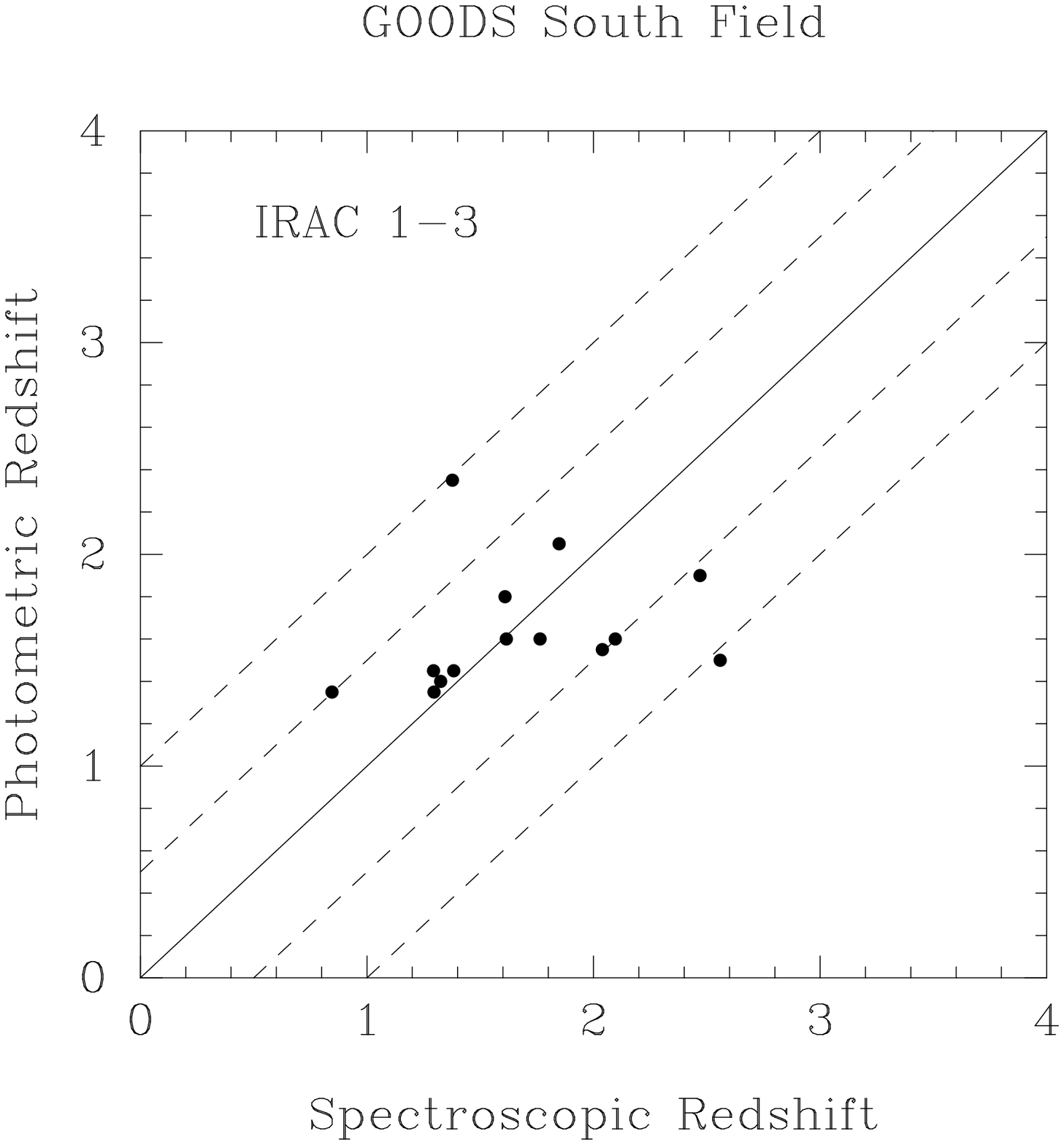}
\caption{\label{fig:no4}Photometric redshift as a function of spectroscopic redshift for galaxies that were fit with $z_{phot}>3$ with all four IRAC bands redone omitting the 8.0\um\ band where photometric uncertainty is highest. Includes 87 galaxies in the North field but only 14 in the South. A $m_{3.6}-m_{8.0}<0.1$ cut was also imposed to eliminate the majority of low redshift outliers.}
\end{figure} 

Iterating the photometry procedure on objects with $z_{phot}>3$ but omitting the 8.0\um\ band added 28 correct galaxies to our $1.5 \leq z_{phot} \leq 2.5$ bin, but also added a fair amount of contamination. This contamnation was reduced by imposing a color restriction of $m_{3.6}-m_{8.0} \leq 0.1$ without affecting the number of correct galaxies. In total, 28 correct galaxies were added to the redshift bin and 17 low redshift outliers, mostly from just below $z_{spec}=1.5$. This increases the completeness fraction in our redshift catalog to $\sim$60\% with a contamination of $\sim$30\%.

To assess the accuracy of our photometric redshifts, we use the normalized median absolute deviation (NMAD), $\sigma_{\Delta{z}/{(1+z_{spec})}}$, where $\Delta{z}$ is defined as the difference between $z_{phot}$ and $z_{spec}$. The NMAD is equal to the standard deviation for a Gaussian distribution, but is less sensitive to outliers than the RMS standard deviation (see, for example, Ilbert \etal 2009). For the galaxies remaining in our spectroscopic catalog after culling based on color, we found the NMAD to be $\sigma_{\Delta{z}/{(1+z_{spec})}} = 0.15$. For comparison, the accuracy of photometric redshifts in the range $1.5 < z < 3$ derived by Ilbert \etal (2009) was $\sigma_{\Delta{z}/{(1+z_{spec})}} = 0.06$. Obviously, much more accurate photometric redshifts are to be expected when using a greater number of bandpasses (30 in the case of Ilbert \etal 2009). However, 0.15 $z_{phot}$ accuracy is sufficient for many applications, such as the creation of luminosity or mass functions at a certain epoch.      

\begin{figure}
\begin{center}
\includegraphics[width=8cm]{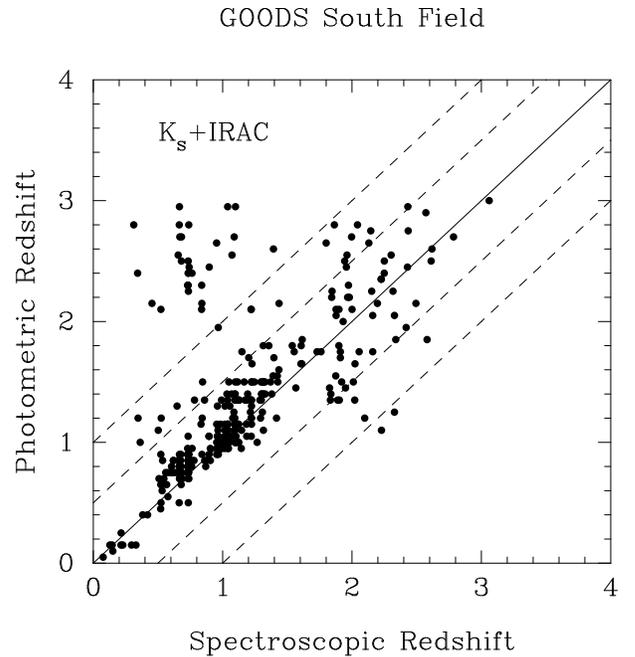}
\caption{\label{fig:K}Photometric redshift as a function of spectroscopic redshift for galaxies in the South field using K-band photometry as well as all four IRAC bands.}
\end{center}
\end{figure}

In the South field, we were able to investigate how the addition of $K$-band photometry can improve results. While the wavelength of the K-band is pushing the limit where changes in the shape of the \bump\ due to model parameters begin to become significant at $z=2$, our limited parameter space should still be acceptable given our modest photometric uncertainties. As can be seen in Figure \ref{fig:K}, including this extra band significantly tightened up the $z_{phot} - z_{spec}$ relationship at lower redshifts. Not surprisingly, the addition of a bluer wavelength did not greatly affect upscattered galaxies at low redshift. Although its addition did lower the number of outliers, it could not compensate for our model templates not properly modeling PAH emission or LIRGS at the IRAC wavelengths. If the \bump\ is to be used to photometrically determine properties of low redshift galaxies (using say $H$, $K$, [3.6], and [4.5]), it is likely that additional templates must be included to fit galaxies that are extremely dusty and infrared luminous. 

From this analysis, we concluded that while IRAC color selection is very efficient at selecting high redshift $z\geq1.3$ galaxies, it is much more difficult to extract any further information from the IRAC photometry. Errors in the 8.0\um\ band can cause the galaxies to be fit to grossly inaccurate photometric redshifts, causing a severe decrease in completeness. Photometry in bands outside the region near the \bump\ would likely help constrain the redshift of these galaxies, but at the cost of having to include more parameters in the models such as reddening and star-formation history.

It is possible that increased signal to noise could make photometric redshifts from solely the IRAC bands much more feasible. JWST will have filters at the same wavelengths as IRAC, but with an extreme increase in sensitivity. Moreover, crowding in the images will be greatly reduced by the superior angular resolution of JWST. These improvements should greatly reduce the photometric scatter in observations, and hence improve the photometric redshift estimation.


\section[Stellar Mass Functions]{Stellar Luminosity/Mass Functions and Stellar Mass Density}
\label{SMF}

In this section we discuss how our catalog of galaxies at $z\sim2$ was used to create a rest-frame $H$-band luminosity function (LF) and stellar mass function (SMF). Working with the IRAC bands greatly simplifies estimating stellar masses for galaxies at this redshift. The rest-frame NIR emission of galaxies arises from comparatively cool stars, which dominate the stellar mass. Furthermore, the NIR spectrum is relatively immune to extinction. Thus, with relatively few model assumptions, we can derive stellar mass estimates. Our stellar mass value for each galaxy simply comes from the stellar mass of the best-fit model multiplied by the scaling factor needed to match the NIR flux of the observed galaxy. To compute our SMF and LF, we first had to correct for both incompleteness and contamination in our redshift catalog.  We divided our corrections into two parts: (1) detection incompleteness, discussed in \S\ \ref{SMF:Veff}, and (2) scatter in our photometric redshifts, which causes both incompleteness and contamination, discussed in \S\ \ref{SMF:Bayes}. 

\subsection{Incompleteness in the Photometric Catalog, The $V_{eff}$ Approach}
\label{SMF:Veff}

We used the effective volume ($V_{eff}$) approach (Steidel \etal 1999, Sawicki \& Thompson 2006) to compute the incompleteness in our photometric catalog due to imperfect object detection efficiency. This approach addresses not only Malmquist bias (brighter galaxies being observable to deeper redshifts), but also the more complicated loss due to varying brightness over different bandpasses. 

We measured the amount of incompleteness in our photometric catalog by implanting simulated galaxies into our images and then attempting to recover them using the same photometry procedure as that used on the original images in \S~\ref{photometry}.
The incompleteness is a function of apparent magnitude, or similarly, a function of stellar mass, with fainter, less-massive galaxies suffering more incompleteness than brighter, more-massive ones. The incompleteness will also be a function of the colors between bands, and hence, the redshifts and intrinsic SEDs of the galaxies. As we discussed before (\S\ \ref{photoz:models}), colors near the \bump\ are not greatly affected by choice of model SED parameters, and we therefore feel justified in simplifying the incompleteness estimation by using only one rest-frame SED to determine the colors of our simulated galaxies. Our simulated galaxy SEDs had an age of 0.5 Gyr with zero extinction, and were redshifted and attenuated using the SEDfit software to give model colors at redshifts between $1.5 \leq z \leq 2.5$ in steps of $\Delta$$z=0.1$. The shape of the artificial objects was assumed to be a point source with the PSF of the detection images, and the shape in each IRAC band was made by convolving the point source with the respective transformation kernel. The simulated objects subsequently had their fluxes scaled to match various apparent magnitudes ($m$) at 4.5\um, with $17 < m < 28$ in steps of $\Delta m=0.5$. Several hundred random locations throughout the images were selected and then the simulated objects were inserted at these locations for each magnitude and redshift in the parameter grid. The fraction of objects recovered forms the completeness function $p(m,z)$, which is the probability that a galaxy of given apparent magnitude (at 4.5\um) and redshift will be present in our photometric catalog.

It is straightforward to convert the recovery fraction to a function of stellar mass,  $\mathcal{M}$, since model mass is determined by the scaling factor needed to create the model apparent magnitude. To derive the absolute magnitude in the rest-frame $H$-band, $M_H$, we used the usual cosmological distance modulus, $DM$, and k-correction, $K$:
\begin{equation}
M_H = m_{\lambda_{obs}} - DM -K.
\end{equation}
This is rewritten as
\begin{eqnarray}
M_H & = & m_{\lambda_{obs}} - 5\log(D_L/10 pc) + 2.5\log(1+z) \nonumber \\
		& & + (m_H - m_{\lambda_{obs}/(1+z)}),
\end{eqnarray}
where $D_L$ is the luminosity distance. The k-correction color between the rest-frame $H$ and the 4.5\um\ filter in the rest-frame of the object is expected to be very small for galaxies at redshifts near 2, and we approximated that term to be zero.

Finally, the effective volume was calculated for each field by integrating the probability function over redshift. For the luminosity function, this is written as
\begin{equation}
\label{eqn:veff}
V_{eff}(M) = A\int_{1.5}^{2.5}{{dV\over{dz}}p(M,z)dz},
\end{equation}
where $dV/dz$ is the comoving volume per square arcminute in redshift slice $dz$ at redshift $z$ and $A$ is the area of the field in arcminutes. The bounds on the integral come from our choice of working in the redshift range $1.5 \leq z \leq 2.5$. Note that, unlike Steidel \etal (1999) or Sawicki \& Thompson (2006), our effective volumes will not approach zero at the integral bounds. In their works, the effective volume corrected not only for detection incompleteness, but for scatter out of their color selection criteria as well. We chose to deal with selection incompleteness in a slightly different manner (\S\ \ref{SMF:Bayes}). The effective volume equation for the stellar mass function is essentially the same as Equation \ref{eqn:veff}, simply replacing $M$ with $\log{(\mathcal{M})}$. The effective volume has a maximum when $p(M,z)=1$ (\ie there is no incompleteness). For our data covering 303.8 arcmin$^2$, this maximum volume works out to be $V_{max} \approx 9.83\times10^5$ Mpc$^3$.

\subsection{Incompleteness and Contamination in our Redshift Catalog, Baysian Inference}
\label{SMF:Bayes}

As discussed in \S\ \ref{photoz:results}, our redshift catalog suffers from incompleteness and contamination from low redshift objects. However, the estimated percentages from our spectroscopic redshift comparison could be biased by incomplete spectroscopy. The small number of spectroscopic redshifts at higher redshifts could lead to gross inaccuracies in our estimates. Our solution was to use the method of Baysian inference described here.

We have created a test, namely, `{\it{does this galaxy lie between redshift 1.5 and 2.5?}}' For simplicity, we will hereafter refer to a galaxy between redshifts 1.5 and 2.5 as being at redshift 2. Let $A$ be the case where a galaxy in our photometric catalog is {\it{actually}} at redshift 2, and let $B$ be the event that our test gives a positive result. We can then define the probability of a true positive, $P(B|A)$, and the probability of a false positive, $P(B|\neg{A})$ where $\neg{A}$ denotes the negation of $A$. If we assume that contamination from high redshifts is negligible, and that the spectroscopic catalog is fairly complete at redshifts lower than 1.5 (both acceptable assumptions), then we can easily estimate the probability of a false positive by our spectroscopic comparison. We could technically use the spectroscopic comparison to find the number of true positives, but there is simply not enough data, especially if we want separate probabilities in each of our magnitude or mass bins. Instead, we ran our simulated galaxies through our redshift fitting procedure and used these to estimate $P(B|A)$.

Once we have estimates of $P(B|A)$ and  $P(B|\neg{A})$, we can use Bayes' Law combined with the Law of Total Probability to estimate what percentage of our positive results are correct, $P(A|B)$, (one minus the percentage of contamination), as well as what percentage of our negative results are actually at redshift 2, $P(A|\neg{B})$ (the incompleteness). Doing so gives equations
\begin{equation}
P(A|B) = {P(B|A)P(A)\over{P(B|A)P(A) + P(B|\neg{A})P(\neg{A})}}
\end{equation}
and
\begin{equation}
P(A|\neg{B}) = {P(\neg{B}|A)P(A)\over{P(\neg{B}|A)P(A) + P(\neg{B}|\neg{A})P(\neg{A})}}.
\end{equation}

Note that $P(\neg{B}|A)$ is simply $1-P(B|A)$ and similarly for $P(\neg{B}|\neg{A})$. The difficulty lies in that we do not know $P(A)$, the probability that a galaxy is actually at redshift 2 in our catalog. We can, however, make an estimate on $P(B)$, the probability of a positive result at any redshift, by simply using our photometric redshift catalog. $P(B)$ is the number of galaxies with $1.5 \leq z_{phot} \leq 2.5$ divided by the total number of galaxies in the catalog. From the Law of Total Probability, $P(A)$ is then given by
\begin{equation}
P(A) = P(A|B)P(B) + P(A|\neg{B})P(\neg{B}).
\end{equation}

These three equations combine to form a cubic equation that can be solved for $P(A)$, two of the solutions always being the trivial cases of all the galaxies actually being at redshift 2 or none of the galaxies actually being at redshift 2. This method also requires that $P(B|A)>P(B|\neg{A})$, or in other words, the test has a higher probability of a positive result when the galaxy is actually at redshift 2.

We found that the percentage of false positives, $P(B|\neg{A})$, and the percentage of positives in general, $P(B)$, did not change significantly with apparent magnitude. We therefore simplified our calculations by holding them as constants, while using our simulated model catalog to calculate $P(B|A)$ for each of our absolute magnitude and mass bins. Our results were that for all masses/magnitudes, the Bayesian contamination was $\sim$10\%. This percentage is half of what was found from the spectroscopic comparison in \S\ \ref{photoz:results}, which indicates that --- as expected ---  there is a bias in spectroscopy at higher redshifts. We also found that the incompleteness was inversely correlated with mass/brightness, such that only 1\% of the galaxies with a negative test result should actually have a positive one for the brightest/most massive bin, but this increased to approximately 10\% in the faintest/least massive bin. The percentages are not exactly the same for luminosity and mass, as there is not a direct conversion between the bins, but the numbers do not differ greatly. In other words, as could be expected, there is a larger scatter in the photometric redshift estimate for fainter objects.  

\subsection{Results}
\label{SMF:Results}

\begin{figure}
\includegraphics[width=8cm]{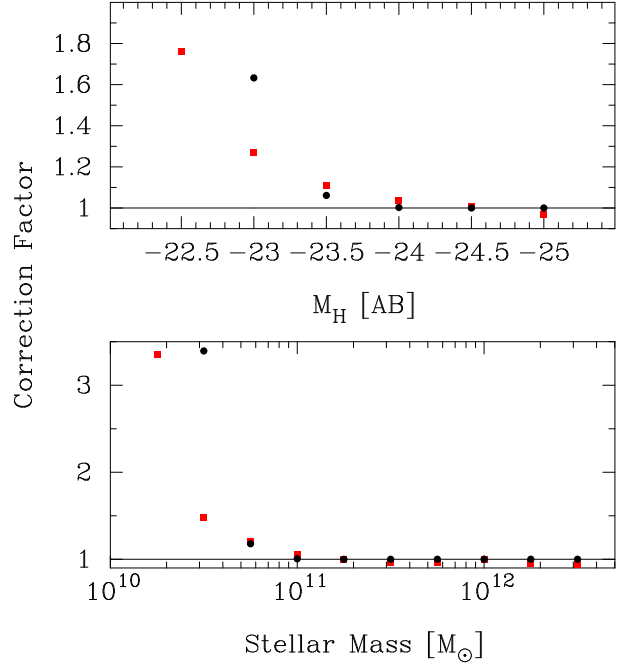}
\caption{\label{fig:correct}Statistical corrections to the number counts in our LF (top panel) and SMF (bottom panel). Black circles show the correction factor for the effective volume (\ie $V_{max}/V_{eff}$) and red squares show the Baysian correction factor for both incompleteness and contamination [$P(A|B) + P(A|\neg{B})(1-P(B))/P(B)$]. When the Baysian correction factor is less than one, contamination dominates over incompleteness in our photometric redshifts, which only happens for the brightest/most massive objects. The $V_{eff}$ correction dominates over the Baysian one at faint magnitudes/low masses. Not pictured in the graphs due to scale restrictions are the $V_{eff}$ correction factors for $M_H = -22.5$ or $\log(\mathcal{M}) = 10.25$.}
\end{figure}

Using the methods above, we corrected our original number counts in each of the bins, $N(M)$ where $M$, depending on context,  represents either the rest-frame $H$-band magnitude ($M_H$) or  the logarithm of the stellar mass ($\log \mathcal{M} $).  The corrected  number count per comoving cubic megaparsec are given by 
\begin{eqnarray}
\phi_{data}(M) & = & [ N(M)P(A|B,M) \\
			   & &  + N(M)P(A|\neg{B},M)(1-P(B))/P(B) ] / V_{eff}, \nonumber
\end{eqnarray}
where the first term on the right-hand side corrects for contamination in our redshift catalog, the second term corrects for incompleteness due to scatter in our photometric redshifts, and dividing by the effective volume, $V_{eff}$, corrects for detection incompleteness. The error bars were determined using Poissonian statistics in the raw number counts, and binomial statistics in the detection counts of simulated galaxies in the $V_{eff}$ calculation. We do not use data points for bins which have an effective volume less than 66\% of the maximum volume, as correction terms would dominate over the data. We found that the Baysian correction for contamination was negligible and the Baysian correction for incompleteness was on par with the $V_{eff}$ correction until the empirical cutoff in the 8.0\um\ bandpass was reached, when the $V_{eff}$ correction began to dominate (see Figure \ref{fig:correct}).

We fit the binned data with the appropriate Schechter (1976) function. The LF is represented as
\begin{eqnarray}
\phi_{model}(M) & = &  \phi^*{\ln(10)\over2.5} \times \left[10^{\left({M^*-M\over2.5}\right)}\right]^{(\alpha+1)} \nonumber \\
				& &   \times \exp\left[-10^{\left({M^*-M\over2.5}\right)}\right],
\end{eqnarray}
and the stellar mass function as
\begin{eqnarray}
\phi_{model}(\log(\mathcal{M})) & = & \phi^*{\ln(10)} \times \left[10^{\left({\log(\mathcal{M})-M^*}\right)}\right]^{(\alpha+1)} \nonumber \\
							& & \times \exp\left[-10^{\left({\log(\mathcal{M})-M^*}\right)}\right].
\end{eqnarray}

\begin{figure}
\includegraphics[width=8cm]{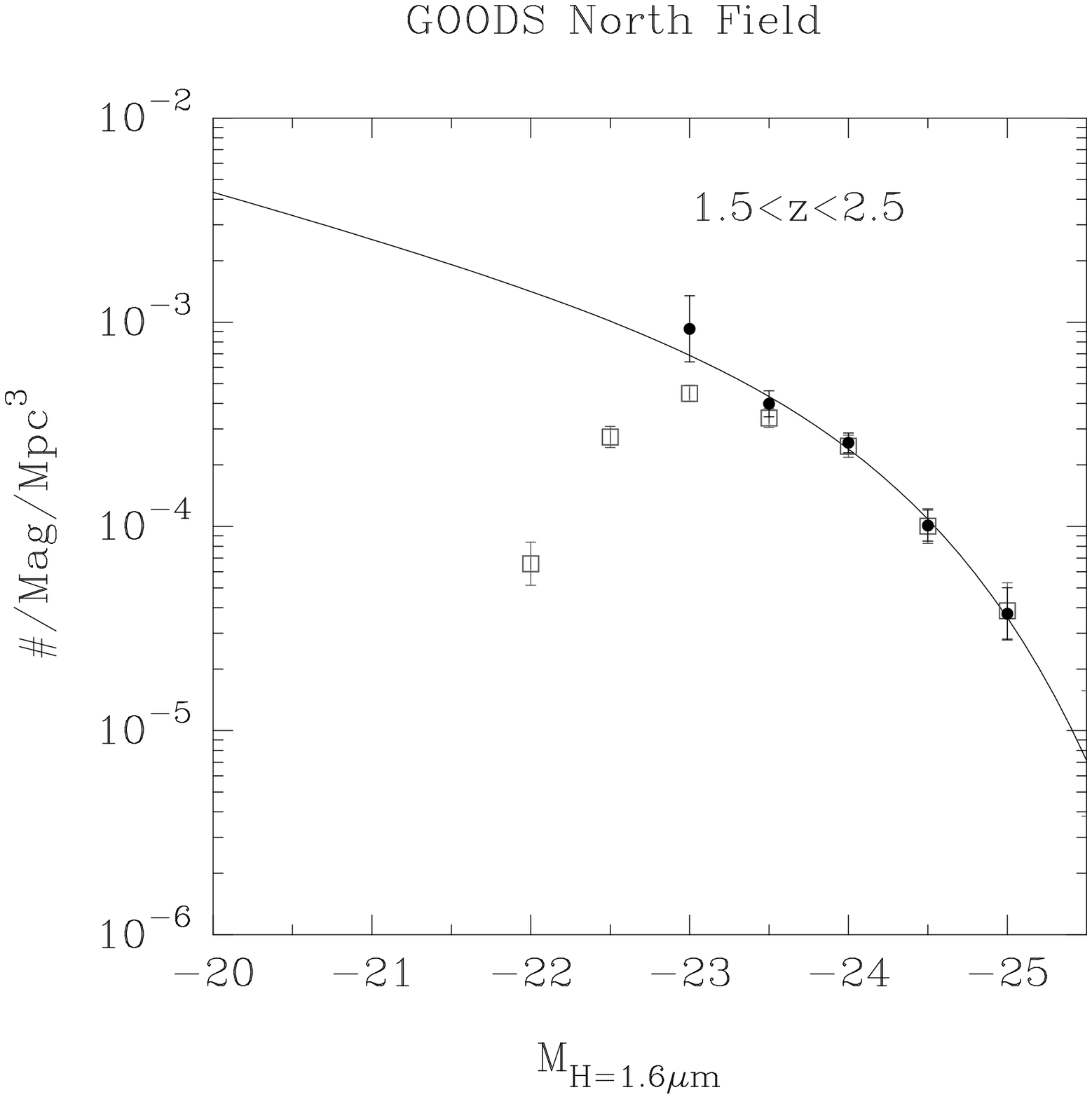}
\includegraphics[width=8cm]{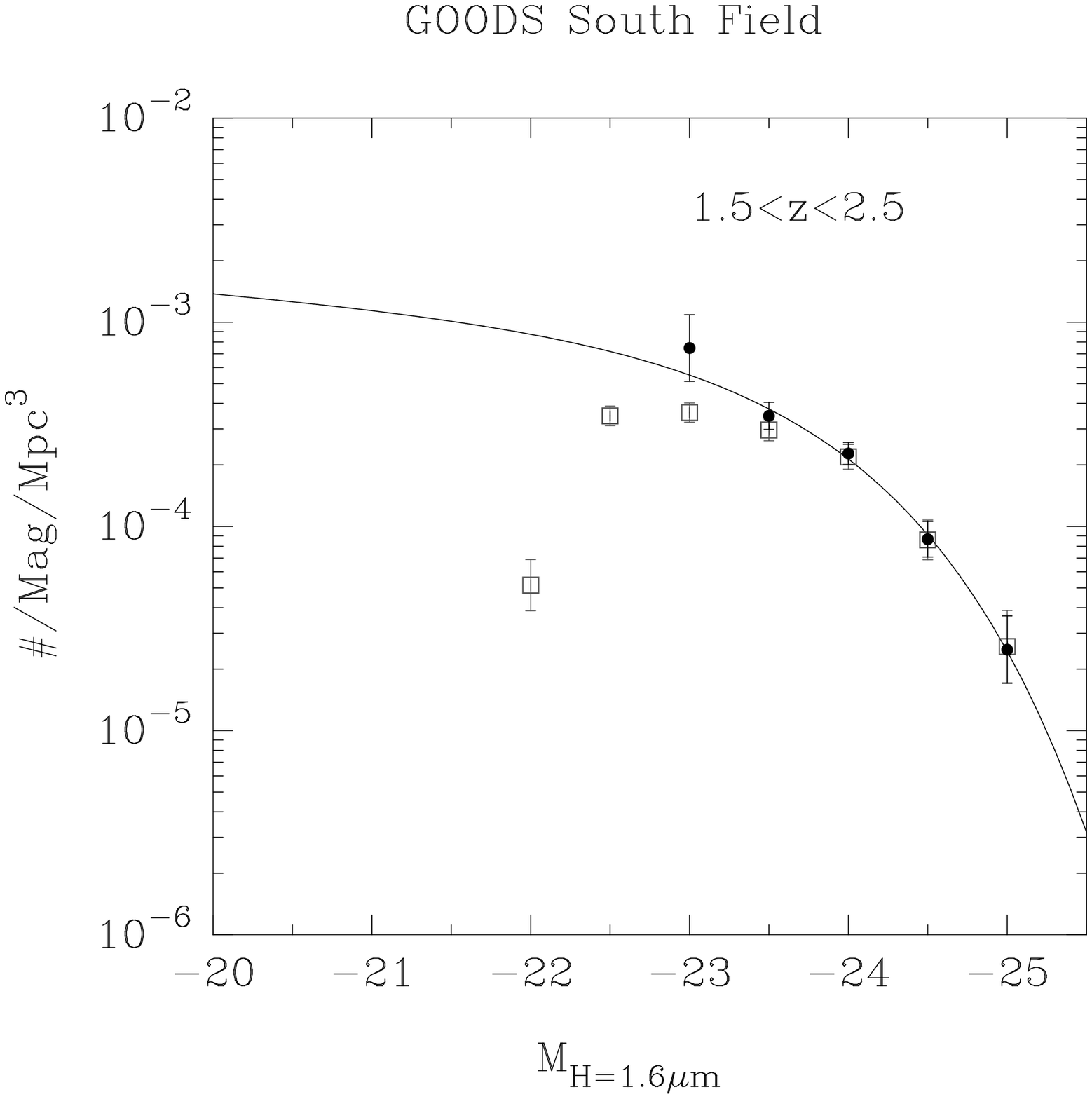}
\caption{\label{fig:LF}Rest-frame H-band luminosity functions for a redshift range of 1.5 $\leq z \leq$ 2.5 in both GOODS fields. Open squares show data scaled by the maximum volume without any correction for incompleteness or contamination from low-$z$ galaxies. Black circles show the corrected data using the 1/$V_{eff}$ and Bayesian inference techniques up to the appropriate completeness level. The solid line shows the best fitting Schechter function.}
\end{figure}

\begin{figure}
\includegraphics[width= 8cm]{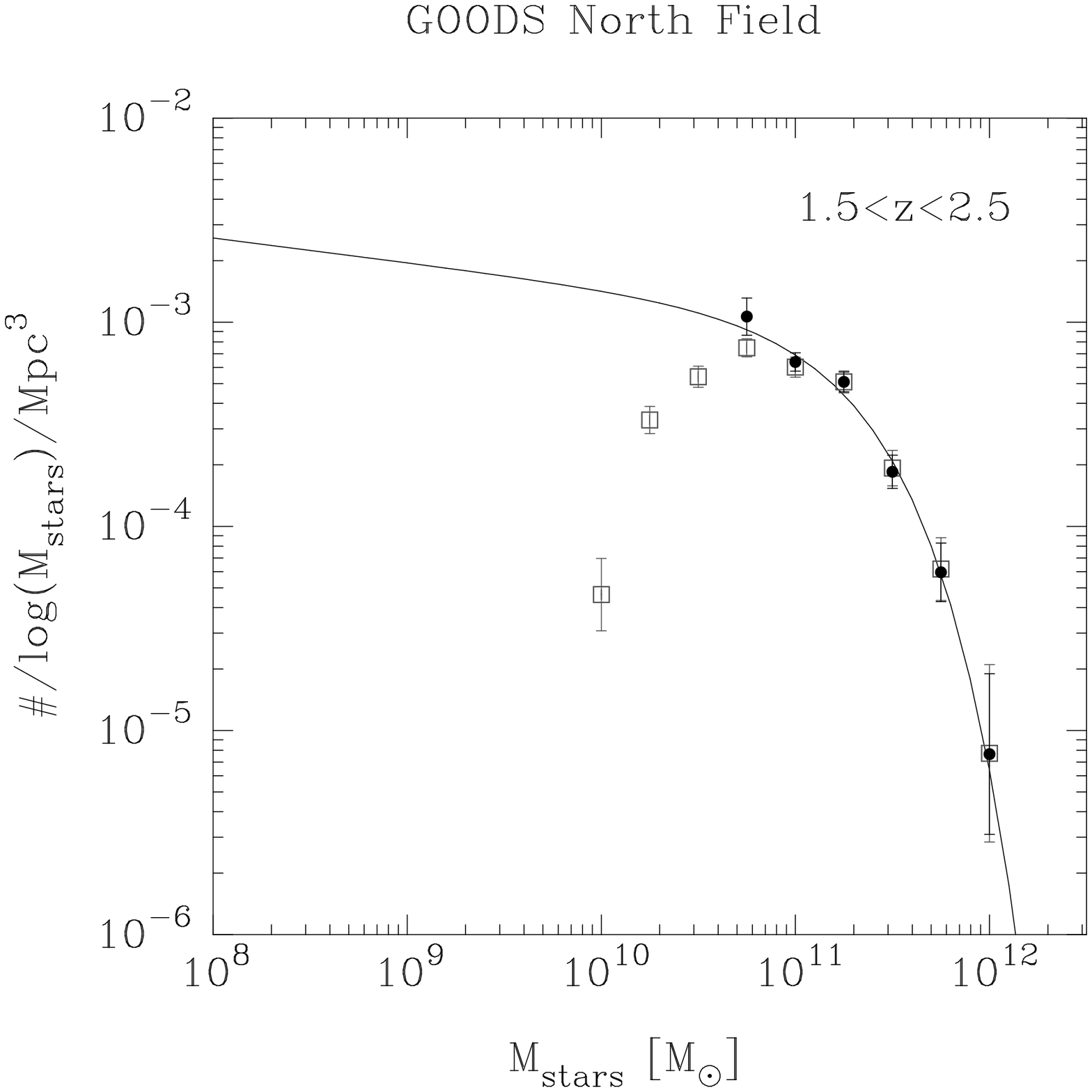}
\includegraphics[width= 8cm]{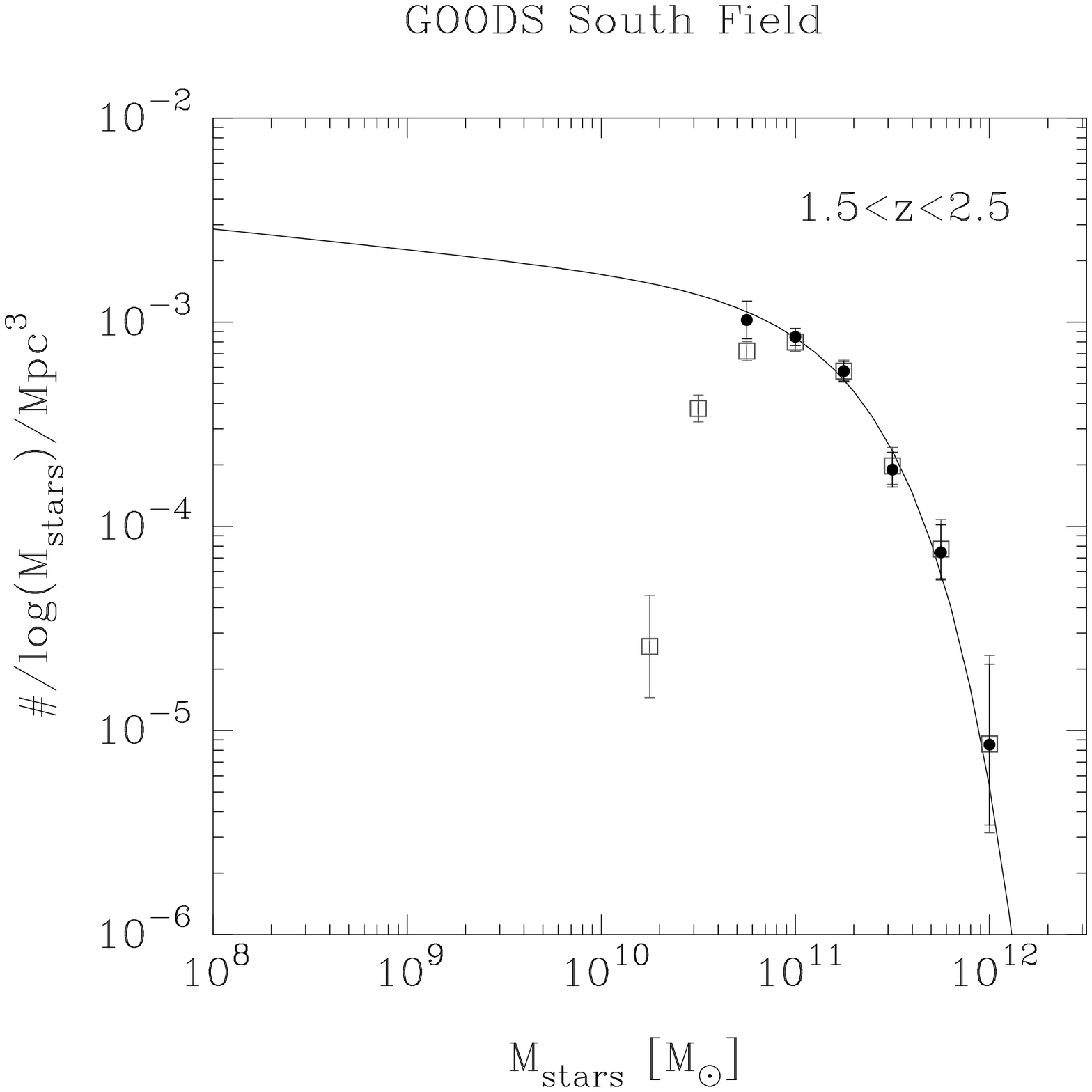}
\caption{\label{fig:MF}Stellar mass functions for a redshift range of 1.5 $\leq z \leq$ 2.5 in both GOODS fields. As in Figure \ref{fig:LF}, black circles and open squares show completeness-corrected and uncorrected points respectively, and the solid line is the best fitting Schechter function to the corrected points.  Masses were determined from the best-fit models and scaling factors found during the SED fitting procedure.}
\end{figure}

We evaluated the best fitting parameters $\phi^*, M^*, \alpha$ using a $\chi^2$ statistic  
\begin{equation}
\chi^2 = \sum_M{\left[{\phi_{data}(M) - \phi_{model}(M)\over\sigma(M)}\right]^2},
\end{equation}
which is linear in $\phi^*$, and so the optimal value of $\phi^*$ is derived by taking $d\chi^2/d\phi$ and setting it equal to zero to yield the equation
\begin{equation}
\phi^* = {\sum_M\hat{\phi}(M)\phi_{data}(M)/\sigma^2(M)\over\sum_M\hat\phi^2(M)/\sigma^2(M)}
\end{equation}
where
\begin{equation}
\hat\phi(M) = {\phi_{model}(M)\over\phi^*},
\end{equation}
(see also Sawicki \& Thompson 2006). Equations 4.10, 4.11, and 4.12 are the same for the SMF, but with $M$ replaced with $\log(\mathcal{M})$. The Schechter function is non-linear in the other two parameters, however, and so we calculated $\chi^2$ values over a grid of parameter values and then searched the grid for the minimum $\chi^2$ value. We adopted those parameters as the best fitting ones and they are listed in Table~1. The data and best fitting functions are plotted in Figures \ref{fig:LF} and \ref{fig:MF}. The separate number counts in each of the two GOODS fields were then added together to create a combined LF and SMF.

\input tab1.tex

\begin{figure}
\includegraphics[width=8cm]{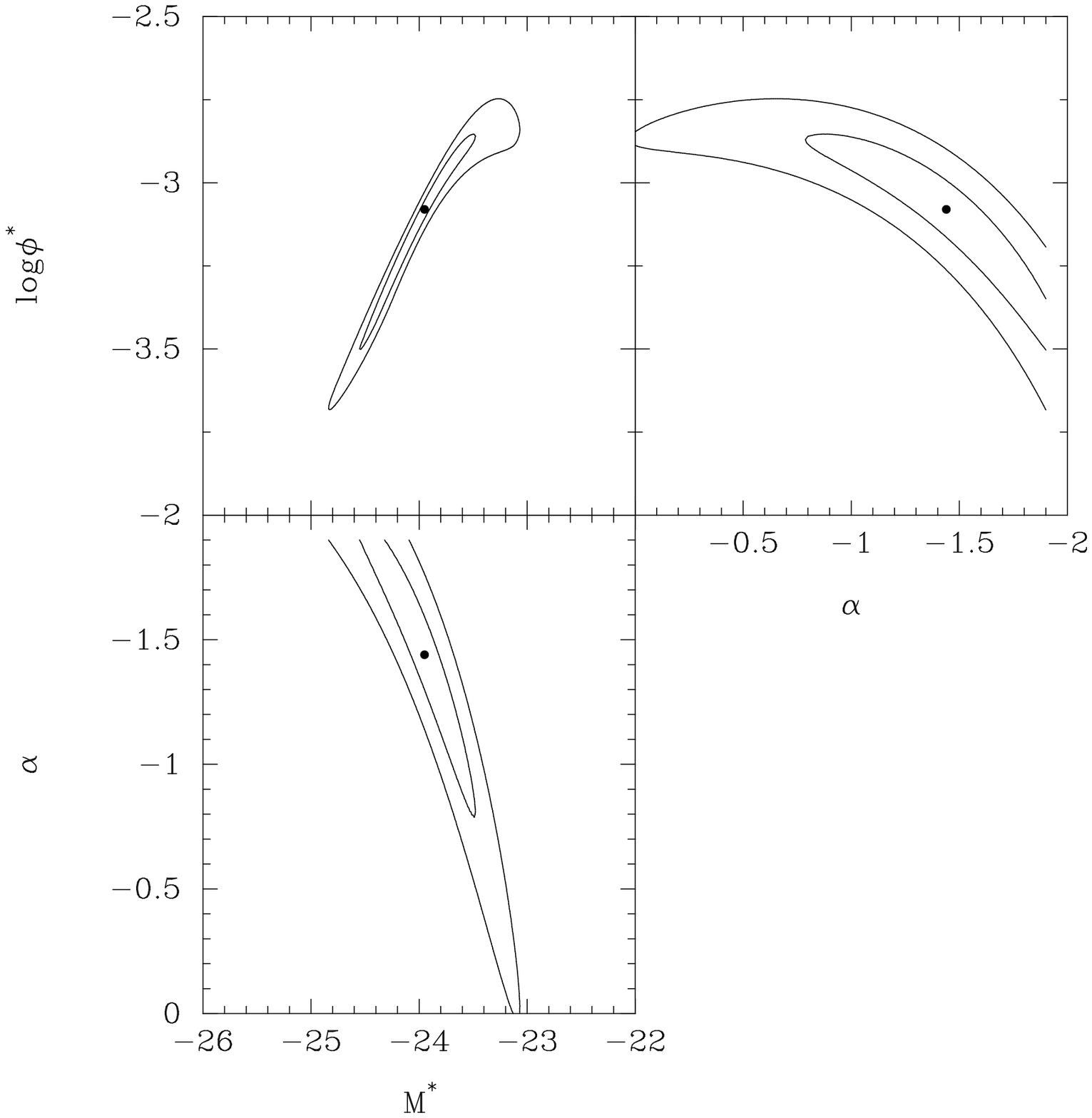}
\includegraphics[width=8cm]{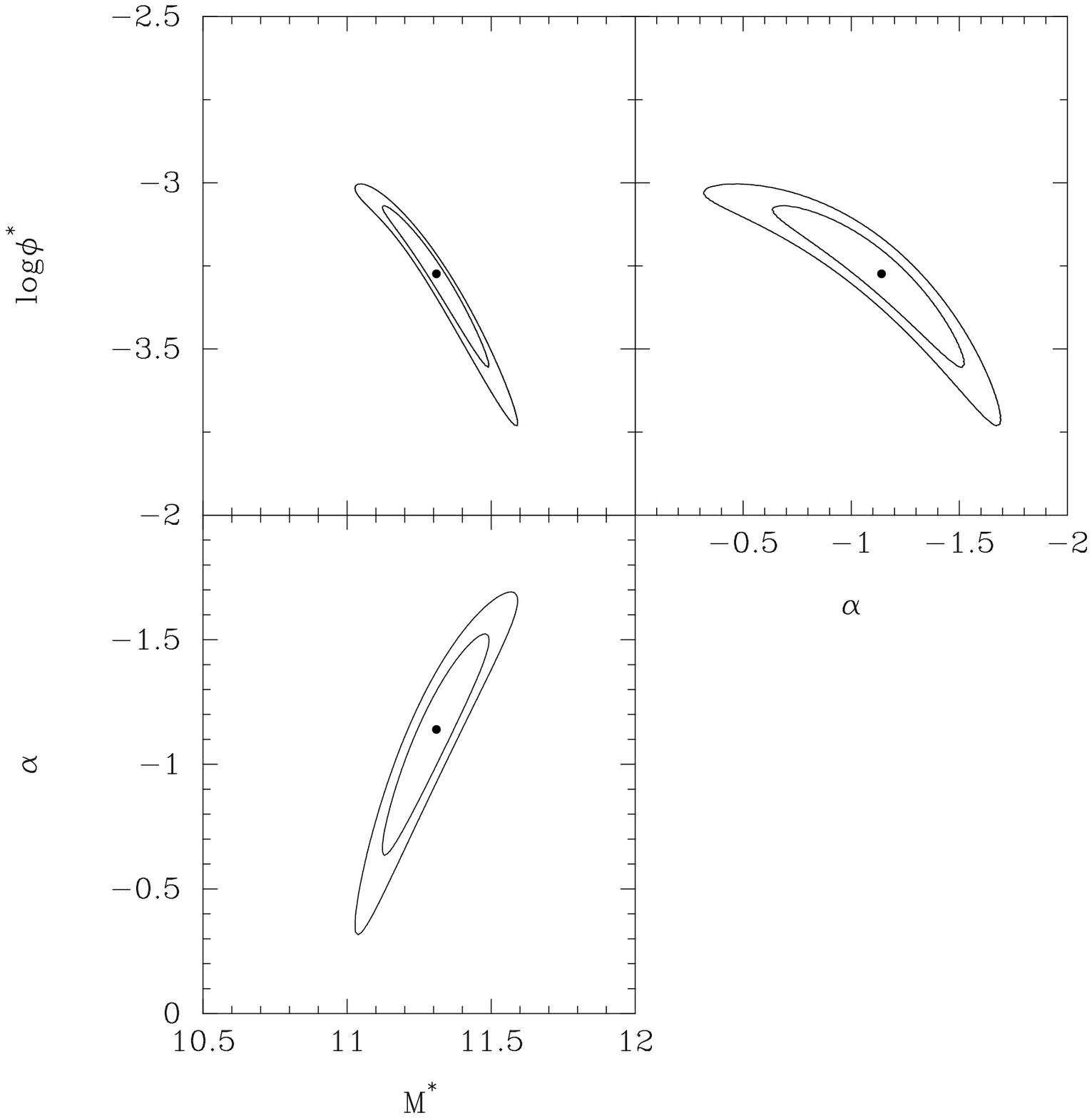}
\caption{\label{fig:contour}1$\sigma$ and 2$\sigma$ confidence intervals for the parameters in the Schechter fits of the combined data of the rest-frame H-band luminosity function (top) and stellar mass function (bottom).}
\end{figure}

Error contours for the best fit parameters were computed by recalculating the best fitting $\phi*, M*$, and $\alpha$, but with values $\phi_{data}(M)$ that have been perturbed randomly according to their standard deviations. We generated 250 perturbed realizations and used their $\chi^2$ value to map out the regions of parameter space that correspond to the best fitting 68.3\% of these realizations. The resulting contours for the combined data are shown in Figure \ref{fig:contour}. Our need for accurate photometry in the 8.0\um\ bandpass severely limits the depth of our data, and results in a poorly constrained faint end of the LFs/SMFs.

\subsection{Discussion}
\label{SMF:Discussion}

\begin{figure}
\includegraphics[width=8cm]{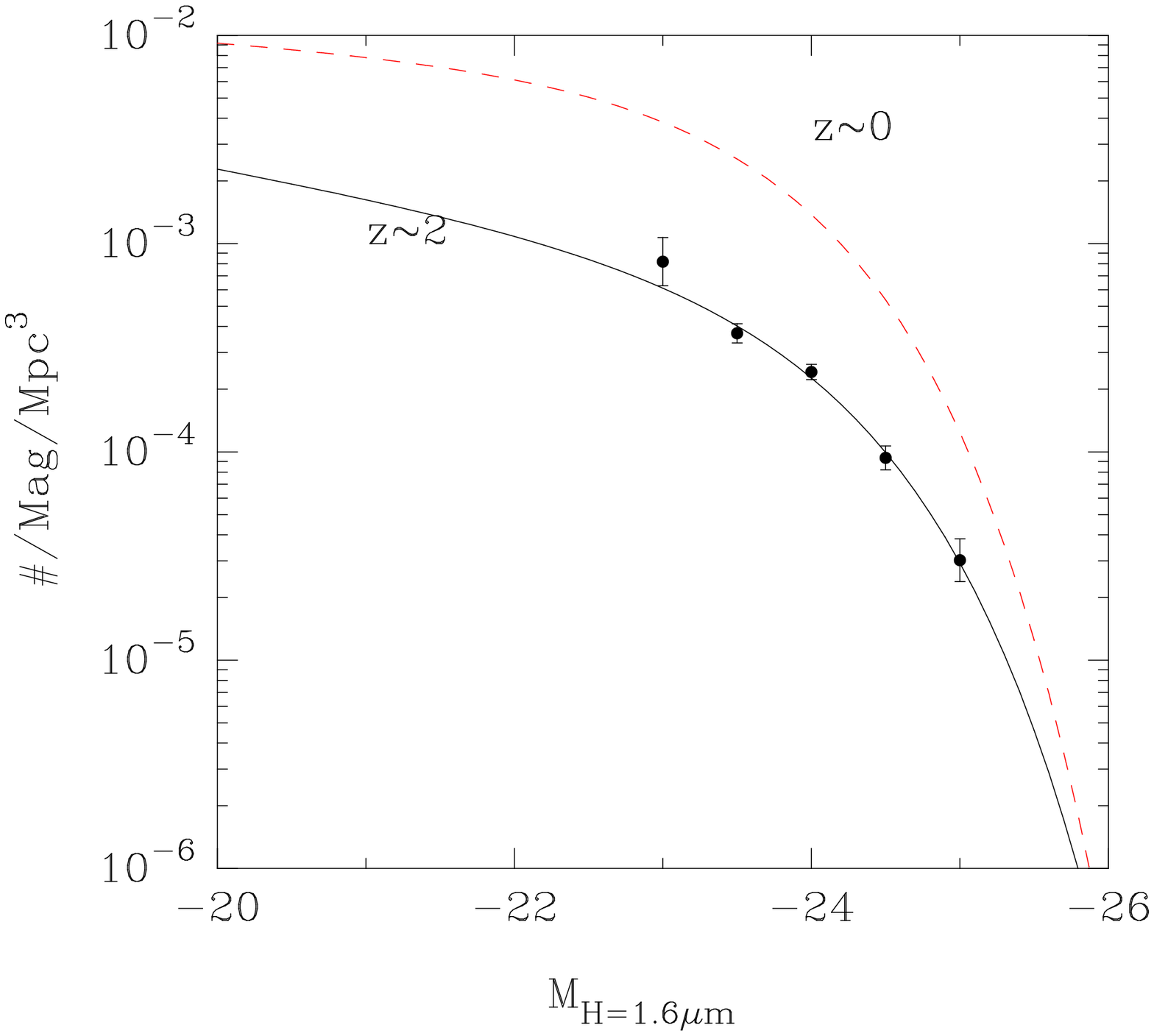}
\caption{\label{fig:LF-comb}Rest-frame $z$$\sim$2 H-band LF obtained by combining data from both the GOODS fields.  The red dashed lines shows the local H-band LF (Jones \etal 2006) for comparison.}
\end{figure}

In Figure \ref{fig:LF-comb} we compare our rest-frame $H$-band LF at $z\sim2$ with $z=0$ results of Jones \etal (2006) to show the evolution of the luminosity function with redshift. Note that our use of an empirical 8.0\um-band cutoff could result in a bias of our LF towards galaxies which are brighter at longer wavelengths. While effective volume corrections (see \S\ \ref{SMF:Veff}) should account for a large portion of galaxies excluded due to a low 8.0\um\ (rest-frame $\sim$2.7\um) flux, these corrections are imperfect, and the model galaxies used in creating them may not fully represent the entire range of galaxy SEDs. Thus, our selection criteria mean that the LF favours galaxies with low levels of extinction, and it should not be treated as a purely $H$-band selected LF.

\begin{figure}
\includegraphics[width=8cm]{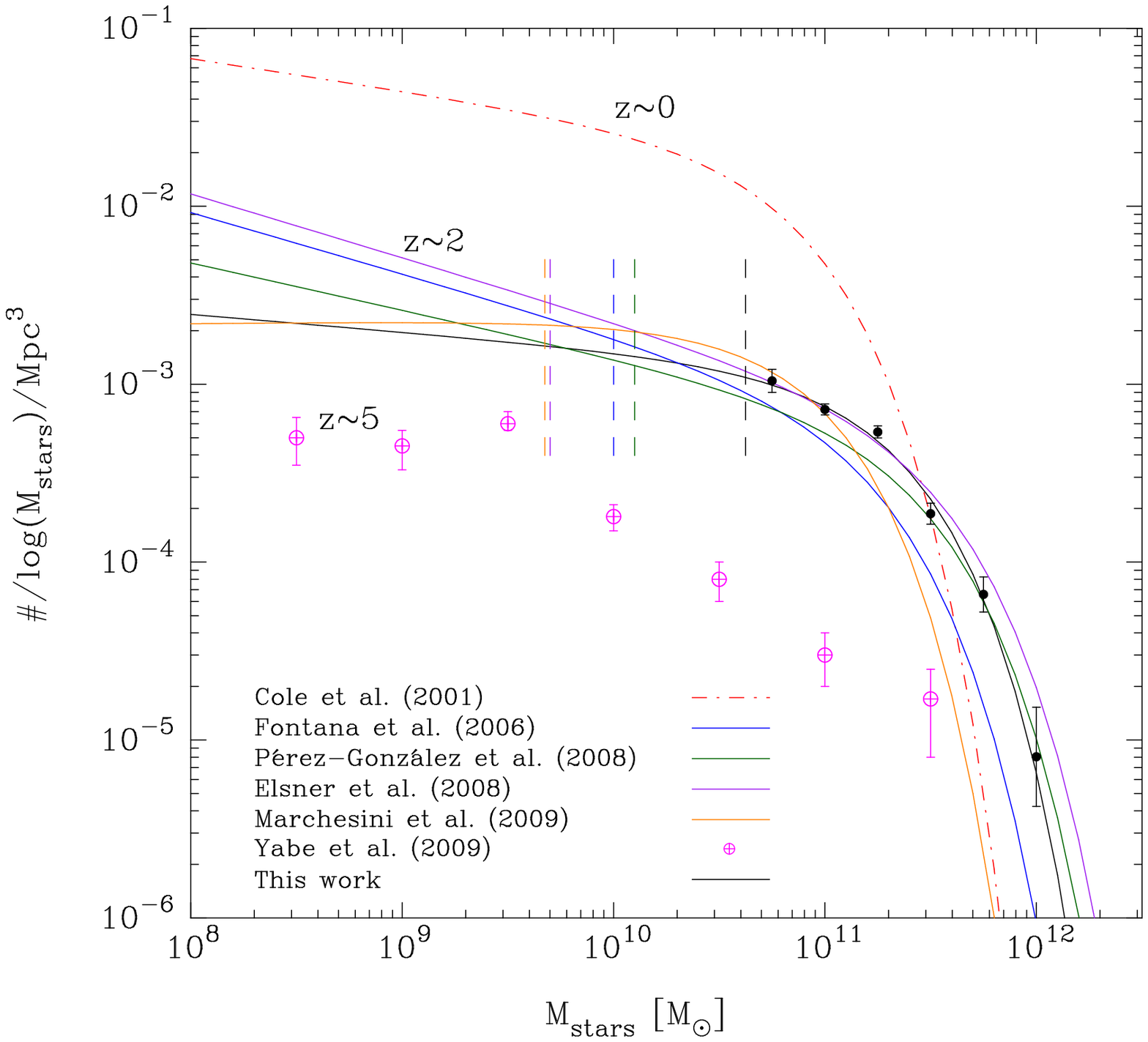}
\caption{\label{fig:MF-comb}Comparison of our combined SMF from both fields with others at the same redshift. The vertical dashed lines show the approximate completeness limits of each function. The dashed red line shows the local SMF (Cole \etal 2001) for reference.}
\end{figure}

In Figure \ref{fig:MF-comb} we compare our stellar mass functions with others. The red dashed line shows the local stellar mass function of Cole \etal (2001) and the crossed circles show LBG (\ie rest-UV and hence star-forming selected) results at $z\sim5$ from Yabe \etal (2009). The other curves show various other best-fit stellar mass functions at approximately the same redshift at $z\sim2$.

\begin{figure}
\includegraphics[width=8cm]{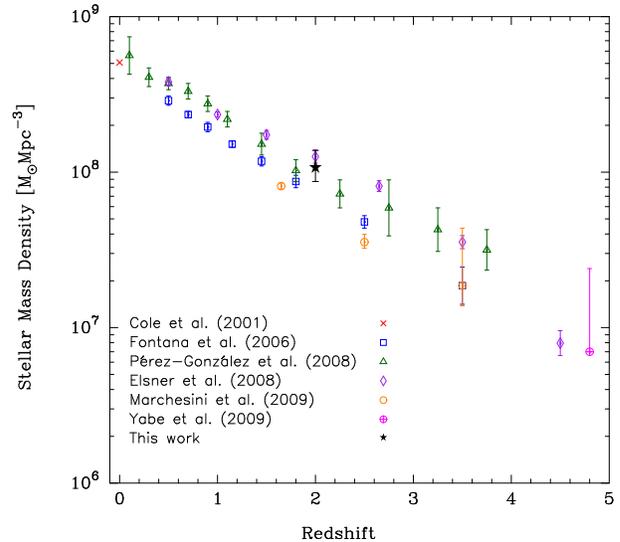}
\caption{\label{fig:smd}Stellar mass density as a function of redshift. This work's result is indicated by the black star, and the work of others is shown as the various hollow symbols. The stellar mass densities have been calculated by integrating the stellar mass functions down to $10^8\Msun$. The lower limit in our data point was taken by summing the stellar masses of all galaxies in our redshift catalog and dividing by the maximum volume between redshifts 1.5 and 2.5, and the upper limit comes from the uncertainty in our Schechter function.}
\end{figure}

In Figure \ref{fig:smd}, we show the integrated stellar mass density of our stellar mass function, as well as the results of other works at different redshifts. Here, we can see the general trend of decreasing mass with increasing lookback time. By redshift 2, the universe had created approximately 1/5 of its stellar mass.

In Figure \ref{fig:MF-comb}, work by Fontana \etal (2006, hereafter F06) and Elsner \etal (2008, hereafter E08), shown as the blue and purple curves respectively, both use the GOODS-MUSIC catalog (Grazian \etal 2006) as the source of their photometry. This catalog comprises multi-wavelength data for 14 847 objects selected in the $z_{850}$ and/or $K_s$ bands, which, at $z=2$, correspond to rest-frame wavelengths of $\sim$0.28 and 0.73\um\ respectively. Colors were measured using a PSF matching technique similar to the one described in \S\ \ref{photometry:crowd} and redshifts determined photometrically using the 14 available bandpasses. Both works use the same spectral synthesis models to estimate stellar masses (Bruzual \& Charlot 2003 with a Salpeter IMF), but F06 use a paramater grid which spans star-formation history, metallicity, age, and extinction, whereas E08 use multi-component models that allow for a recent star-burst phase, but restricted their models to solar metallicity. Both works used the Calzetti (2000) extinction law. The difference between the blue and purple curves in Figure \ref{fig:MF-comb} is thus a good representation of the uncertainties that systematics in differing modeling procedures can produce, without effects from different selection criteria.

An IRAC 3.6 and IRAC 4.5 selected stellar mass function (P{\'e}rez-Gonz{\'a}lez \etal 2008, hereafter PG08) is shown by the green curve in Figure \ref{fig:MF-comb}. This sample consists of 27 899 objects in the Hubble Deep Field North (HDF-N), Chandra Deep Field South (CDF-S) and the Lockman Hole field (LHF). Aperture photometry was measured in the IRAC bands and a correction factor based on empirical PSFs was applied to obtain a total magnitude. Model templates were generated using the PEGASE code (Fioc \& Rocca-Volmerange 1997) and spanned a parameter space of star-formation history, metallicity, age, extinction, and allowed for a second component of a recent instantaneous burst of star formation. The attenuation at any wavelength was calculated using the Charlot \& Fall (2000) recipe.

The orange curve shows a recent SMF from Marchesini \etal (2009, hereafter M09) that was made using a $K$-selected sample constructed with the Multi-wavelength Survey by Yale-Chile (MUSYC; Gawiser \etal 2006), the Faint Infrared Extragalactic Survey (FIRES; Franx \etal 2003), and the GOODS-CDFS FIREWORKS catalog (Wuyts \etal 2008). Fluxes were measured using aperture photometry where the aperture's size and shape were optimized based on simple criteria such as the galaxy's isophotal area and whether or not the galaxy was blended. Photometric redshifts were derived using a non-negative linear combination of PEGASE model templates, while stellar masses were derived assuming BC03 models of solar metalicity, a Kroupa (2001) IMF, and the Calzetti extinction law. 

\begin{figure}
\includegraphics[width=8cm]{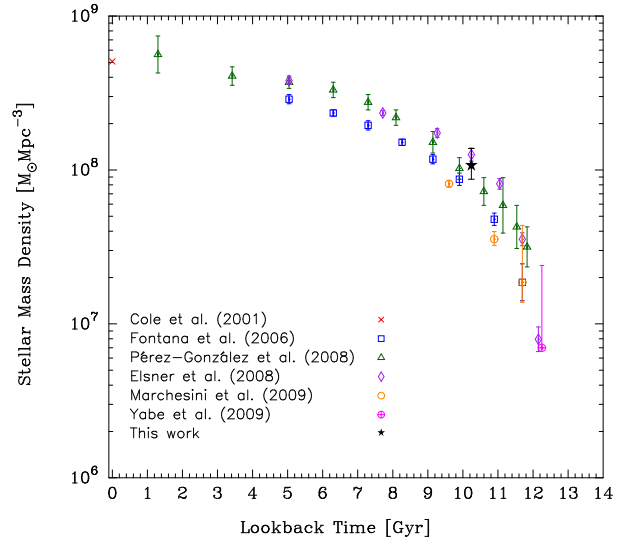}
\caption{\label{fig:lookback}Same as Figure \ref{fig:smd}, but this time showing the stellar mass density as a function of lookback time.}
\end{figure}

We found  good agreement between our results and the others. The differences in the low-mass slope are well within the large uncertainty of our error contours, but we caution that deeper data, especially in the 5.8 and 8.0\um\ bands, is needed to properly compare the low-mass end. Discrepancies at the most massive bins are most likely due to small numbers at these masses as well as cosmic variance. The agreement of our stellar mass function with those made using a far greater number of bands leaves us confident that the \bump\ can be used to accurately estimate both the redshift and stellar mass of a galaxy. 

Comparing the IRAC-selected SMFs (this work, PG08) with those selected at shorter wavelenghts, we do not see a significant discrepancy. This suggests that there is not a large amount of stellar mass at this redshift that is missing in surveys selected at $K_s$ (rest-frame $\sim0.73$\um) or bluer wavelengths.

Examining evolutionary trends in the SMF, we see results favoring the ``downsizing" scenario for galaxy formation. The massive end of the SMF seems to be already in place by $z=2$. Since then, the creation of stars had to have happened predominantly in less massive ($<10^{11}\Msun$) galaxies in order to match the local stellar mass function. The faint end slope of the SMF is not well constrained at high redshift in any of the works and pushing observations to fainter limits is of great interest as the majority of stellar mass in the universe lies in low-mass galaxies. Future observations with JWST will probe an unsurpassed depth at high redshift with a far superior angular resolution compared with IRAC. Rest-frame NIR selection techniques, such as those that use the \bump\, when used with JWST will be able to provide a great amount of information regarding the evolution of low-mass galaxies.


\section{Conclusions}
\label{conclusions}

In this work we used Spitzer/IRAC imaging to explore the feasibility and limitations of using the \bump\ to select high-redshift galaxies, to estimate their redshifts, and to study the global properties of the population, namely its rest-frame H-band luminosity function and its mass function.  Our main conclusions are as follows.

\begin{enumerate}
\item The \bump\ is feasible as a means of selecting galaxies and determining their redshifts photometrically. Using only IRAC's 3.6, 4.5, and 8.0\um\ filters, galaxies with redshifts greater than 1.3 can be selected on a color-color diagram using the criteria $m_{3.6}-m_{4.5} \geq 0.12(m_{3.6}-m_{8.0} )-0.07$. This method of selection is very complete (greater than 90\% complete as estimated by spectroscopic redshifts), but with a good deal of contamination from low redshift galaxies ($\sim 33\%$). One can lower the contamination rate at the expense of completeness by increasing the intercept of this selection criteria. For example, the criteria $m_{3.6}-m_{4.5} \geq 0.12(m_{3.6}-m_{8.0} ) + 0.02$ has $\sim16\%$ low-z contaminates, but only $\sim75\%$ completeness.

\item Information from all four IRAC bands can be used to obtain photometric redshifts fairly accurately in the range $1.3 < z < 3$, although with a large amount of scatter, mostly due to poor S/N in the 8.0\um\ bandpass. We stress here the importance of accurate photometry in the fourth ``anchor" bandpass in constraining redshifts of galaxies. Even with information from all four bands, there is still a great deal of contamination from upscattered low redshift galaxies in the photometric redshift catalog, but the majority of this can be dealt with by using the color selection criteria discussed above, or more simply only including galaxies with an $m_{3.6} - m_{4.5}$ color less than -0.1. We found that the contamination and incompleteness can be dealt with effectively using the statistical method of Bayesian inference.

\item The addition of bluer bands such as $K_s$ and possibly $H$ can increase the feasible range of this technique down to redshift zero. However, for best results at $z<1$, a broader range of model templates must be included to account for LIRGs and PAH emission. Similarly, including redder bands could theoretically be used to push the limits of this technique to earlier redshifts. However, doing so would require a great deal of sensitivity at these wavelengths, which is only expected to be possible with future instruments such as JWST.

\item In the case of galaxies whose SEDs are dominated by the very youngest stellar populations (less than 0.01 Gyr, which should be a very small percentage of galaxies) it is impossible to constrain the redshift, as the power law slope of the spectrum is degenerate with redshift. Any galaxies with a best-fit model ages less than this age should have their results treated with a great deal of skepticism. 

\item The shape of the \bump\ is very robust, and we found that photometric redshifts in the range $1.3 < z < 3$ were not greatly affected by choices in model parameters. This robustness means that the \bump\ cannot tell much about these parameters, but it also means that only a small parameter space is required for model template fitting and reduces the chances for degeneracies in the models. There is, however, a degeneracy between age and mass in the models, and as such, input model templates should be limited to realistic ages to avoid systematic biases in mass estimation. 

\item By simply estimating the stellar mass of each galaxy based on the scale factor required to match the model flux with the observed flux of the \bump\, we generated a stellar mass function for galaxies at redshift $\sim2$. Comparing our results with others that used a far greater number of bandpasses (10 or more) and a much larger model parameter space showed good agreement to within the uncertainties. This leaves us confident that the \bump\ can be used to efficiently and effectively estimate redshifts and stellar masses of galaxies. We found no evidence for a significant amount of stellar mass missing from surveys selected in bluer ($z$ and $K$) bandpasses. 

\item Our study is consistent with the ``downsizing" scenario of the evolution of cosmic stellar mass density. The massive end of the SMF was already in place by $z\sim2$ and since then, star-formation must have happened primarily in lower-mass galaxies. Our findings are not consistent with a simple hierarchical scenario. This suggests that there must be some mechanism that shuts down star-formation in the most massive galaxies at high redshift.
\end{enumerate}

The ability to select and study galaxies from the rest-frame NIR {\it{without any information from shorter wavelengths}} will be a valuable tool for JWST. The usable wavelength range of JWST (0.6--27\um) will make it impossible for current selection techniques (Lyman Break, $BzK$) to select moderate redshift galaxies, whereas the last four broadband filters of JWST's NIRcam instrument (central wavelengths of 2.0, 2.77, 3.56, and 4.44\um) will allow selection using the \bump\ at redshifts less than $\sim1.5$. The first two broadbands of the MIRI instrument (central wavelengths of  5.6 and 7.7\um) can take the place of IRAC's 5.8 and 8.0\um\ bands to extend this selection to $z=3$, and redder bands could extend even further in redshift. The far greater resolution and sensitivity of JWST compared to Spitzer/IRAC should greatly increase the accuracy of photometric redshifts, and resolve many of the issues originating from difficulty in obtaining quality IRAC photometry. The unprecedented depth of JWST will place tight constraints on the faint end properties of luminosity functions and stellar mass functions. The \bump\ is well poised to tell us a great deal of information about the galaxies observed with JWST, without the need to survey fields in a large number of bandpasses.



\acknowledgments

\vspace{5mm}

We are grateful to the GOODS team, those at the ESO/GOODS project, and P. Capak \etal for providing all the publicly available data that made this work possible.  This work was supported by grants from the Natural Sciences and Engineering Council of Canada and the Canadian Space Agency. Parts of the analysis presented here made use of the Perl Data Language (PDL) that has been developed by K.\ Glazebrook, J.\ Brinchmann, J.\ Carney, C.\ DeForest, D.\ Hunt, T.\ Jenness, T.\ Luka, R.\ Schwebel, and C.\ Soeller and which can be obtained from http://pdl.perl.org.  PDL provides a high-level numerical functionality for the perl Scripting language (Glazebrook \& Economou, 1997).






\appendix
\setcounter{figure}{1}

\section{PHOTOMETRIC TECHNIQUE}

The essential assumption of our photometric procedure is, the photometric procedure assumes that galaxies that are confused in the low resolution, longer wavelength image (hereafter the measure image) are resolved in a higher resolution, shorter wavelength image (hereafter the detection image). The process of using the detection image to constrain photometry in the measure image is as follows: 1) Each galaxy in the detection image is convolved with a transformation kernel in order to match the PSF of the measure image. 2) The convolved galaxies are normalized to unit flux, yielding a model profile for each galaxy in the measure image. 3) The normalized model profiles are each scaled simultaneously to obtain a best-fit to the measure image. 

In more detail, to find the scaling factors, we constructed a $\chi^2$ statistic of the form
\begin{equation}\label{eq.a1}
\label{chisq}
\chi^2 = \sum_{xy} {(I(x,y) - B - \sum_{i=1}^Nf_iP_{i}(x,y))^2\over\sigma_{RMS}^2(x,y)},
\end{equation}
where $I(x,y)$ is the value of the $x$th and $y$th pixel in the measure image, $B$ is an estimate of the background throughout that image, $\sigma_{RMS}(x,y)$ a root mean square (RMS) map of the measure image, and $P_i(x,y)$  is the model profile for each galaxy $i$ through $N$ (created using the method described above). The sum is over all the pixels $x$ and $y$. Minimizing this statistic with respect to each free parameter $f_i$ (which in physical terms represents the flux of each galaxy) leads to a system of equations of the form
\begin{equation}
\mathbf{Af} = \mathbf{b}
\end{equation}
where the boldface indicates that these are matrices. The components of these matrices are given by
\begin{equation}
A_{ij} = \sum_{x,y}{P_i(x,y)P_j(x,y)\over\sigma_{RMS}^2(x,y)}
\end{equation}
and
\begin{equation}\label{eq.a4}
b_i = \sum_{x,y}{P_i(x,y)[I(x,y)-B]\over\sigma_{RMS}^2(x,y)}
\end{equation}
and {\bf{f}} is a column vector containing the various flux scalings $f_i$.

The matrix {\bf{A}} is very sparse, having non-zero components only where model profiles overlap in the measure image (\ie where galaxies are blended), and can be inverted easily using standard numerical techniques. The scaling factor (or flux) for each galaxy can then be solved for, and the uncertainty for each flux is the square root of the diagonal terms of the inverse of matrix {\bf{A}}. Given the zero-point magnitude ($ZP$) of the measure image, one can convert these scaling factors to apparent magnitudes ($m_i$) by 
\begin{equation}
m_i = -2.5\log(f_i) + ZP
\end{equation}

\subsubsection{Assumptions}
\label{photometry:crowd:assumptions}

This technique has many underlying assumptions that, if incorrect, could affect the quality of the resulting photometry, and it is important to understand the limitations of this algorithm. In this section we discuss these assumptions in detail and how they may or may not be addressed if invalid.

Galaxies must be isolated in the detection image in order to be deblended in the measure image. Even with extremely high resolution data, some galaxies will still overlap due to superposition along the line of sight, or simply due to real physical proximity. Overlap in the detection image means that one can only get a flux estimate for both of these galaxies together. However, if the two overlapping galaxies have vastly different colors between the detection and measure images, the combined flux estimate in the measure image can still be in error. The color difference between bands leads to a shift in the location of the brightness peak, which can result in a poor model fit. Fortunately, this effect should be small given a detection image with high spatial resolution.

Along the same lines, it is assumed that the morphology of a galaxy is the same in both the detection and measure images. This assumption may not be valid for real galaxies, whose morphologies may vary at different wavelengths because of the prominence of different processes (\eg star formation, thermal dust emission, flux from old stars dominating at longer wavelengths, \etc). These effects could lead to vastly different spatial profiles and we originally found that residuals often had a prominent peak at the center of each galaxy, most likely due to the prominence of the bulge in the IR (see also De Santis \etal 2007, Laidler \etal 2007). We found that this residual can be adequately dealt with by adding a second, ``Mexican Hat" component to the model profiles, as discussed in the Procedure section of this Appendix.

The RMS uncertainty used in Equation \ref{chisq} comes entirely from the measure image, which is only valid if the uncertainty in the measure image is much greater than that of the detection image. In our case, the low signal to noise ratio of the IRAC images means that uncertainty in the detection image can safely be ignored (the RMS in the IRAC images is typically 8-10 times larger than the $z$ or $K_s$ images). However, if this assumption were not valid, a total RMS uncertainty could be created by adding the detection and measure uncertainties in quadrature.

Another important underlying assumption is that galaxies present in the measure image also have counterparts in the detection image. Depending on the relative depth of each image and the colors of the observed objects, this may not always be true.  As long as the galaxy missing from the detection image is isolated in the measure image, this will have no effect on the fluxes of other galaxies. If, however, the missing galaxy is blended with a neighbouring galaxy in the measure image, the neighbour galaxy's flux estimate will be overestimated as the algorithm tries to compensate for the light added from the galaxy not present in the detection image. This effect is of great importance in our case, as we would like to have a catalog unbiased by the effect of differing stellar populations. Galaxies which are ``red and dead" (\ie passive or quiescent), are extremely dusty, or are at very high redshift all have a large amount of near-IR and IR flux and will hence be present in the IRAC bands, but have very little blue flux, and could therefore be missed in our detection images, most notably in the $z$-band. As described in the Photometry section of this Appendix,
we correct for this by inserting simulated objects into the detection images at the proper locations as an {\it{ex post facto}} prior. Note that it is acceptable for an object to be present in the detection image but missing in the measure image, as this will yield a best-fit scaling factor of approximately zero, and one can still ascertain an upper limit on the object's flux. Conversely, if the detection image is too deep, \ie contains a very large number of objects compared to the measure image, then the flux fits become degenerate, and one cannot trust the results. 

It is imperative that both the detection and measure image are properly aligned astrometrically. A shift in the brightness peak by more than a few pixels between the model and measure image will have drastic detrimental effects on the best-fit flux. Although this issue could be addressed by adding additional degrees of freedom and allowing the model to shift in pixel space, we do not investigate this solution at this time. Instead, we visually confirmed that our images appeared to be well aligned by inspecting small galaxies with diameters less than three pixels across and verifying that they were at the same position in all images.

In generating the model profiles, it is assumed that an accurate transformation kernel has been obtained to change the galaxies from the PSF of the detection image to that of the measure image. Obtaining such a kernel is non-trivial and we discuss our method in the Photometry section of this Appendix.
The Spitzer PSF varies across the field of view, and to handle this complication, we calculate kernels on image sections that are 2\arcmin$\times$2\arcmin, which are small enough to ensure kernel uniformity. 

It is also important to obtain accurate background estimates for galaxies in both images, as this will affect the resulting scale factor. This effect is particularly prominent for fainter galaxies. Details of our background estimation are discussed in the Photometry section of this Appendix.

Finally, one should note that it is possible for the algorithm to assign unrealistic negative fluxes. This usually occurs with faint (S/N $\lesssim$ 3) galaxies around brighter objects as the algorithm attempts to artificially compensate for a poor fit to the bright object (possibly due to an imperfect transformation kernel). Although the fitted flux for the bright object will not be greatly affected, one should assign an upper limit to the fainter galaxies in post-processing. We did not include these objects in our final catalog.

We created software, called FOZZY, in order to carry out the crowded field photometry, along with the companion software KERMIT, which finds transformation kernels between images as described above. These codes were written using Perl and PerlDL (Glazebrook \& Economou 1997).  The underlying principles of this software are explained in the following sections.  In the overlap region of the GOODS fields, the rotation of 180$^\circ$ between epochs causes the PSF to be oriented differently in each epoch of observations. We chose to work with each epoch separately, instead of trying to combine the images, which would only lead to a more complicated transformation kernel.

\subsubsection{Kernel Generation}
\label{photometry:crowd:kernel}

Obtaining an accurate transformation kernel is very important, and here we outline the procedure we followed. First, the images were broken up into several overlapping sub-images in order to account for any variation of the PSF across the field of view (which is prominent in the IRAC images).  Next, we followed the procedure set out by Alard \& Lupton (1998, see also Alard 2000). Briefly, the kernel is assumed to be a linear combination of Gaussians of differing variances multiplied with polynomials: 
\begin{equation}
\label{ker}
K = \sum_{q} a_qe^{-(x^2+y^2)/2\sigma_q^2}x^{m_q}y^{n_q},
\end{equation}
where $q$ has been chosen as the summation variable to avoid confusion with previous equations, and the degree of the polynomials is limited for each variance to some arbitrary degree $D$ such that $0 < m+n \leq D$, with $m, n$ being positive integers. We have chosen variances of $\sigma_q=$1, 3,and 9 pixels with polynomial degrees of 6, 4 and 2 respectively. If we let the Gaussian-polynomial component be represented by $k_q$, then Equation \ref{ker} can be abbreviated to
\begin{equation}
K = \sum_{q} a_qk_q
\end{equation}
and the kernel can then be determined through the use of the $\chi^2$ statistic
\begin{equation}
\label{kerchi}
\chi^2 = \sum_{xy} {(I - \sum_{q}a_q(R \otimes k_{q}))^2\over\sigma_{RMS}^2}
\end{equation}
where $R \otimes k_q$ represents the detection image, $R$, convolved with $q$th gaussian-polynomial, and $\sigma$ is the RMS uncertainty in the measure image.

Differences in background levels between the two images can be fitted simultaneously by assuming the background can be represented by a linear combination of polynomials less than some degree (in our case, $0 < m_r + n_r \leq 3$). This modifies Equation \ref{kerchi} to be
\begin{equation}
\chi^2 = \sum_{xy} {(I - \sum_{q}a_q(R \otimes k_{q}) + \sum_{r}b_rx^{m_r}y^{n_r})^2\over\sigma_{RMS}^2}.
\end{equation}
The system of equations generated by minimizing this statistic with respect to the parameters $a_q$ and $b_r$ can be solved in a manner similar to that of Eq.~\ref{eq.a1} -- \ref{eq.a4}.

\begin{figure}
\begin{center}
\includegraphics[width=0.522\textwidth]{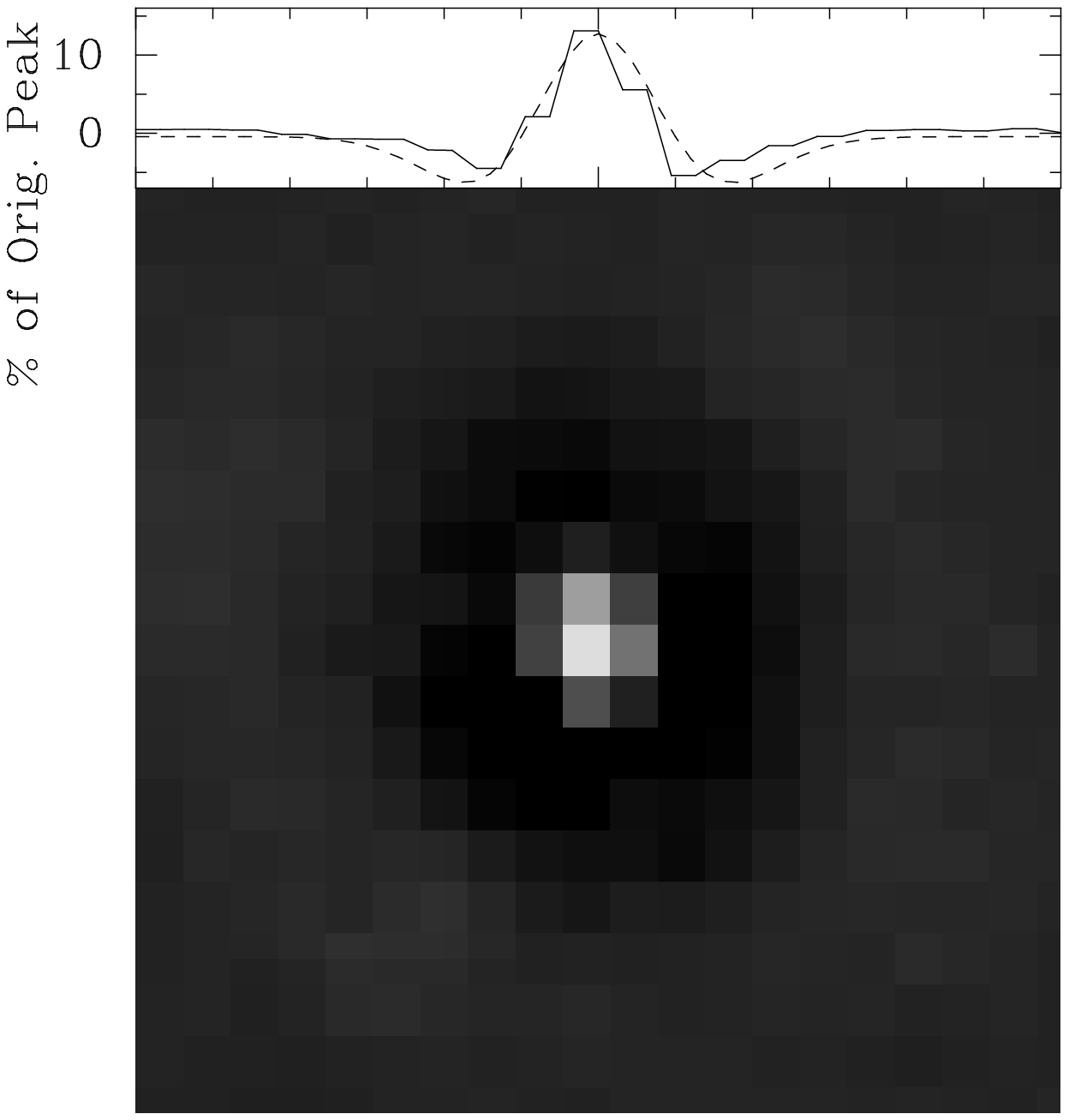}
\includegraphics[width=0.458\textwidth]{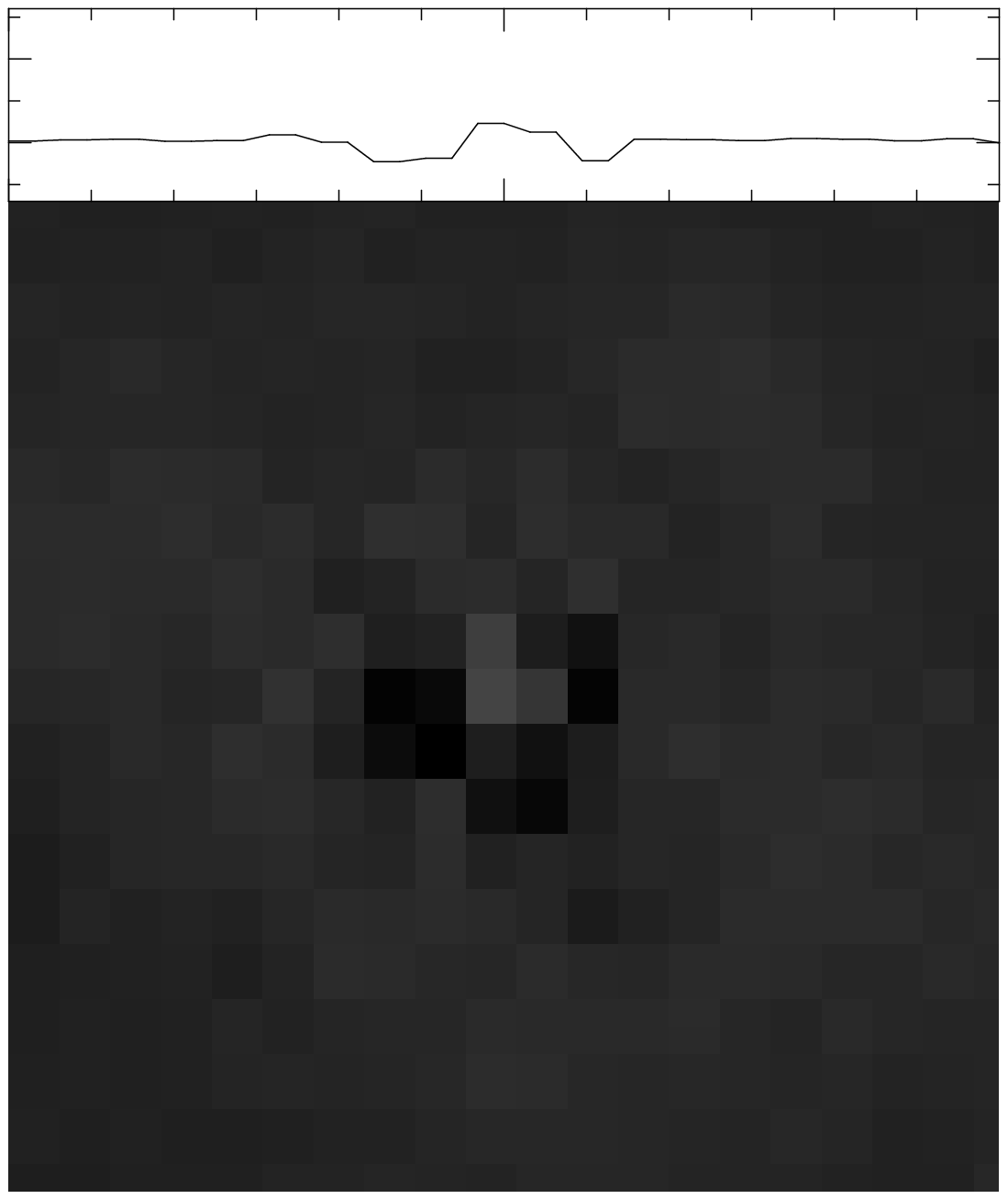}
\caption{\label{fig:badres}a) The left hand image shows a sample residual without the MHF component included in the model. Notice the bright peak surrounded by an over-subtracted area. Shown at top is the percentage of the original peak flux of the galaxy of a slice through the center of the residual (solid line). The bright spot in the center contains a significant percentage (nearly 15\%) of the original peak flux. A MHF of variance 1 is overplotted to show the resemblance (dashed line). b) The right hand figure is the same, except showing the residual when the photometry procedure is done including a MHF component in the models. The residual is now much closer to the level of the noise, and is always less than 5\% of the original peak pixel.}
\end{center}
\end{figure}

 As explained above, residuals of galaxies after subtracting their scaled models often have bright peaks at their center surrounded by over-subtracted regions as shown in Figure \ref{fig:badres} a). To combat this effect, we added an extra component to our fitting procedure of the model profiles convolved with a Mexican Hat Function (MHF), again normalized to unit flux. The formula for the MHF is given by
\begin{equation}
Q_i(x,y) = {1\over\sqrt{2\pi}}(2-x^2-y^2)e^{-{1\over2}(x^2+y^2)}.
\end{equation}

The addition of the MHF component modifies Equation \ref{chisq} to be 
\begin{equation}
\chi^2 = \sum_{xy} {(I(x,y) - B - \sum_{i=1}^N(f_iP_{i}(x,y) +g_iQ_i(x,y)))^2\over\sigma^2(x,y)}
\end{equation}
where $Q_i(x,y)$ is the MHF model component and $g_i$ are the scale factors for that component, and the total flux for a galaxy would be $f+g$. It is clear that this does not alter the solution method described above, but simply doubles the number of free parameters. We found that, in all cases, the addition of this component greatly reduced the overall $\chi^2$ value of the best fit, and, based on simulations, improved magnitude estimates by $\sim$0.1 mag. A sample residual with the MHF component included in the models is shown in Figure \ref{fig:badres} b).

\subsubsection{Photometry Procedure}
\label{photometry:crowd:procedure}

\begin{figure}
\begin{center}
\includegraphics[width=8cm]{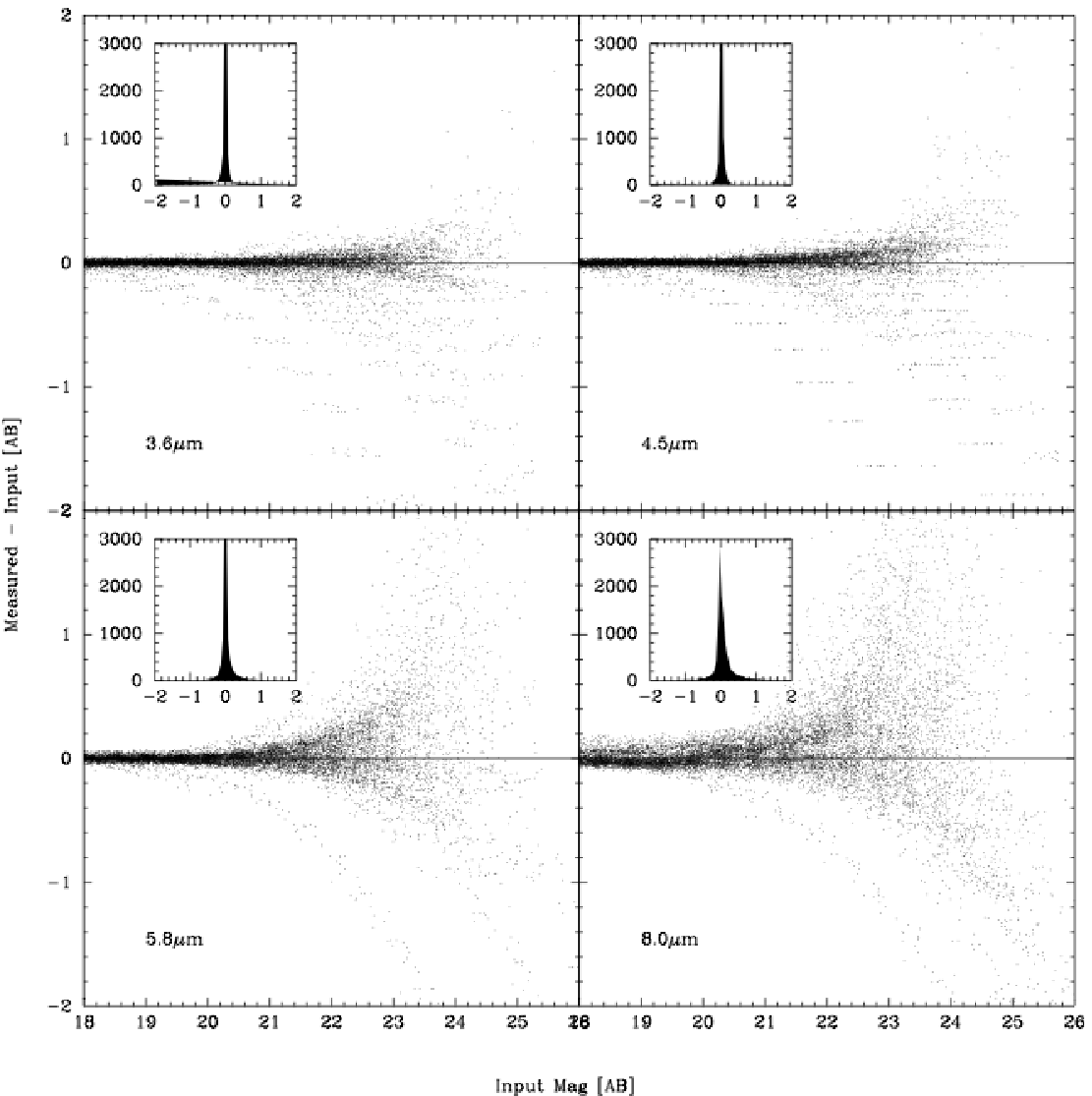}
\caption{\label{fig:magdiff}Difference in magnitude between simulated galaxy input magnitudes and those found by our photometry program (FOZZY) as a function of input magnitude. The inset panels show a binned histogram of number counts in bins of 0.05 mag. Branches diverging from the zero line at lower magnitudes are the result of simulated galaxies placed directly on top of other real galaxies already present in the images.}
\end{center}
\end{figure}

Galaxies were detected in the $z$ (South) or $K_s$ (North) band (detection) images using SExtractor (Bertin and Arnouts 1996),  a software package used in extragalactic astronomy for object detection and photometry. Using this software, we generated a catalog of positions and local background estimates, as well as a segmentation map. Along with its other products, SExtractor creates a segmentation map, which is an image where the pixels attributed to an object have been given values equal to the object's catalog ID number, and pixels not assigned to any object have a value of zero. SExtractor was then used in dual-image mode on each of the IRAC images to determine the local background around each object at these wavelengths. Dual-image mode is a setting that allows SExtractor to detect objects and define photometric apertures in one image, but take all measurements in a different image. This mode allows for easy correlation between objects with measurements in multiple images or bandpasses. The segmentation image was then used to extract galaxies from the detection image as a starting point in generating the model profiles.

As stated above, there is some concern that some galaxies may be bright in the IRAC band passes, but very faint a lower wavelengths and hence missed in the detection image. To counteract this, we ran SExtractor again (in single-image mode) on the 4.5\um\ image and correlated the positions in this catalog with those in the detection catalog using a search radius of 0.9 arcseconds. The number of galaxies detected at 4.5\um\ but not present in the $z$-band was approximately 10\% of the galaxies detected in both bands. This percentage increased to $\sim$30\% in the shallower $K_s$-band detection image. We accounted for the missed galaxies by inserting simulated objects of shape equivalent to the detection image's PSF into the detection image at the position given by the 4.5\um\ catalog. Note that the brightness of the simulated galaxy was arbitrary, but inconsequential as the resultant model profile was subsequently normalized. The requirement that the galaxy was detectable at 4.5\um\ makes our catalog an IRAC-2 selected catalog.


\begin{figure}
\begin{center}
\includegraphics[width=8cm]{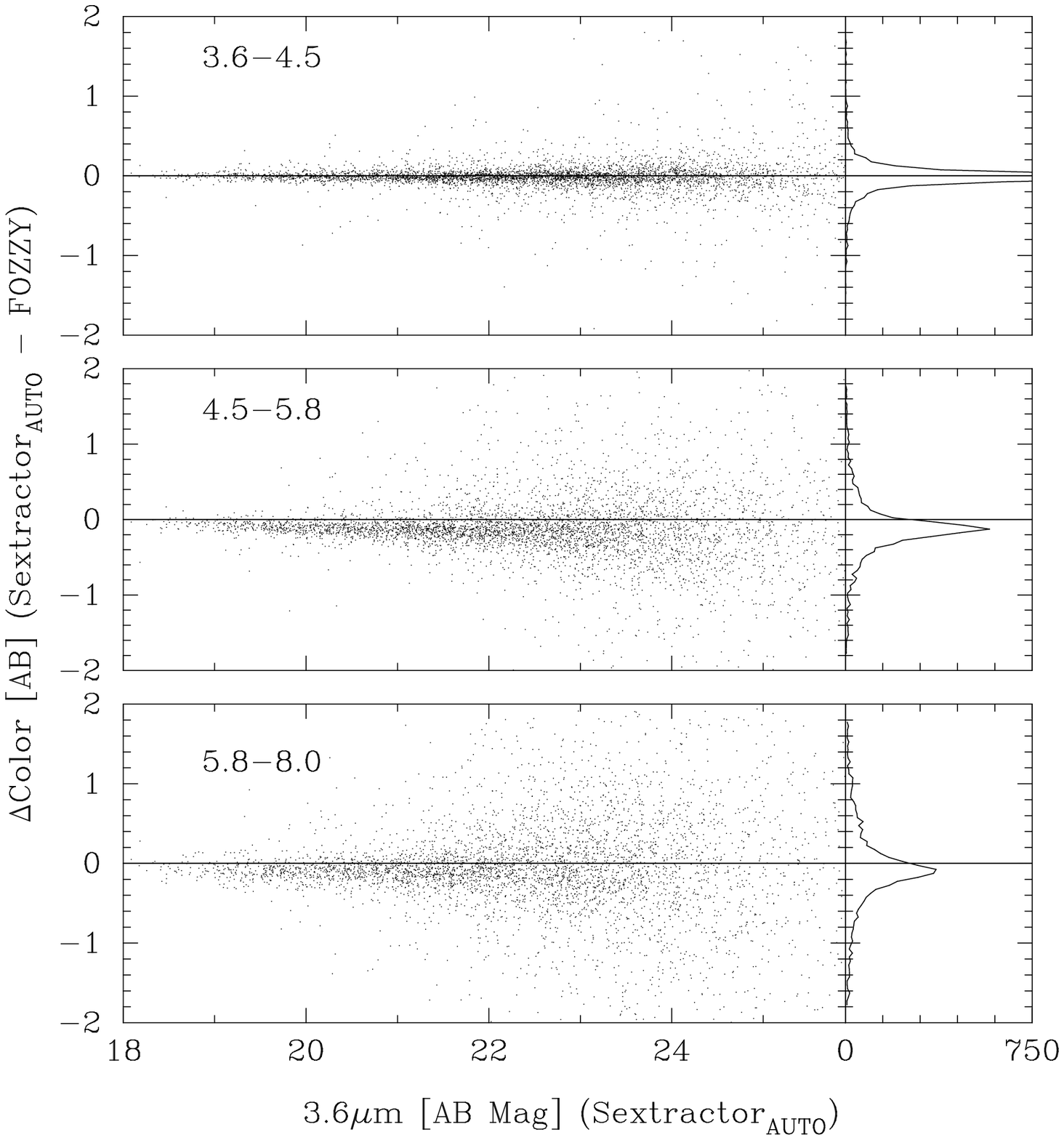}
\caption{\label{fig:testfozzy}Color Differences between SExtractor's MAG-AUTO aperture and our photometry program (FOZZY) for galaxies flagged as isolated by SExtractor (\ie their flux should not be contaminated by neighboring galaxies). The offset from zero at the longer wavelengths is due the aperture from dual-image mode in SExtractor being too small. The right hand panels show binned histograms of number counts in bins of 0.05 mag.}
\end{center}
\end{figure}

 \begin{figure}
 \begin{center}
\includegraphics[width=8cm]{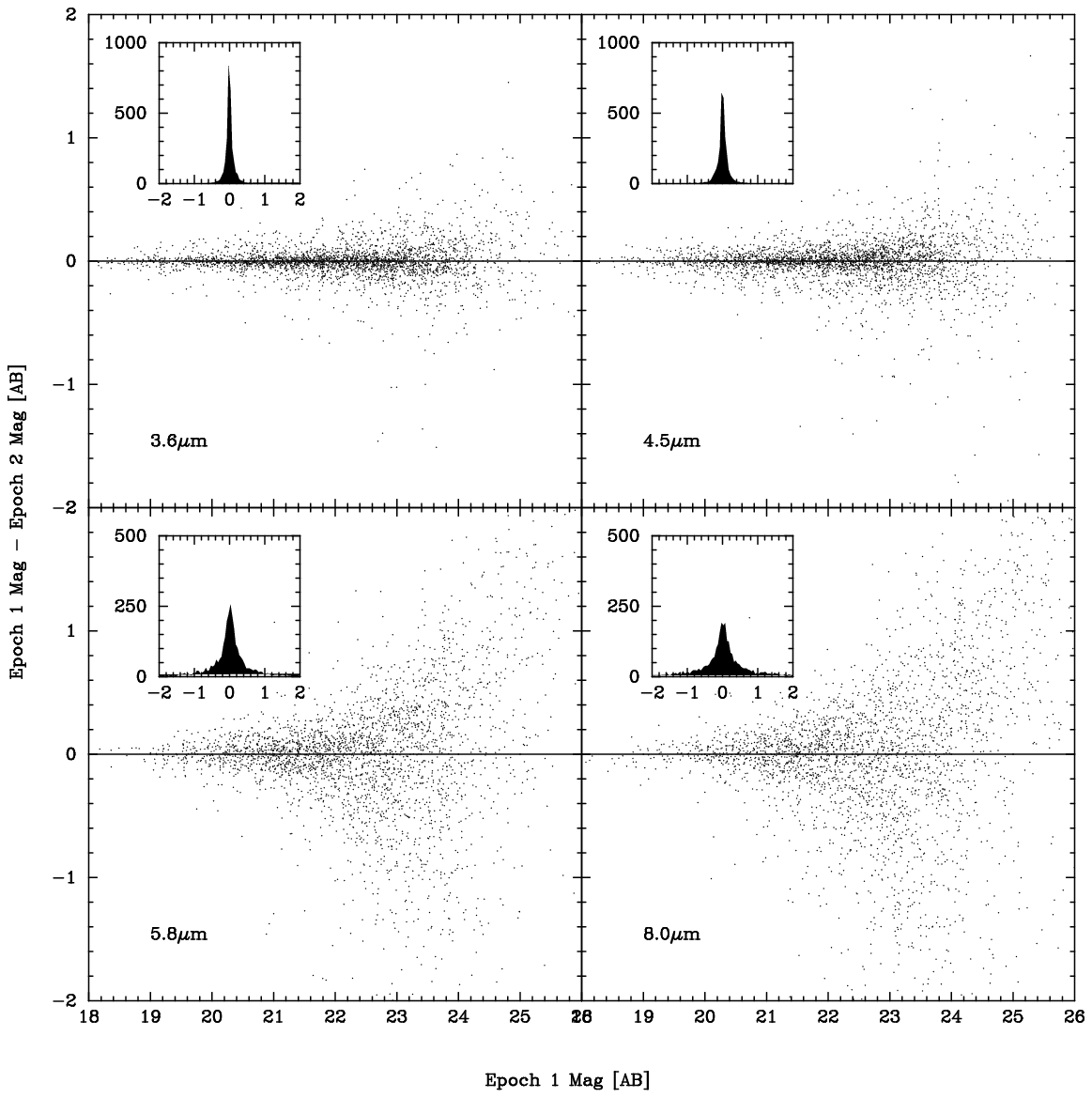}
\caption{\label{fig:overlap}Magnitude differences between the two observational epochs for galaxies located in the overlap region versus magnitude in the arbitrarily chosen first epoch. This is probably the best representation of the typical amount of sky noise or random errors in each bandpass. The inset panels show a binned histogram of number counts in bins of 0.05 mag. }
\end{center}
\end{figure}

In order to test that our procedure produced accurate photometry, we conducted Monte Carlo simulations by inserting simulated galaxies of varying brightness into the images in random positions, which were then put through our photometry program. The shapes of the simulated galaxies were originally equivalent to the PSF of the detection image, but were convolved with the transformation kernels for each of the IRAC bands in order to match the PSF in each image. Although a point source is not realistic, the large PSF of the IRAC images leave no need for accurate spatial resolution in the simulated galaxies. 

Figure \ref{fig:magdiff} shows the difference between the measured and input magnitudes versus input magnitudes. We found good agreement in all bands, although accuracy diminished with increasing wavelength, most likely due to the much lower signal to noise ratio (S/N) in the 5.8 and 8.0 bandpasses. Unlike Grazian \etal 2006, we found no need to dilate the segmentation map in order to account for missed flux at the edges of galaxies. Indeed, we found this was difficult to implement well, as dilating the detection image leads to galaxies bleeding in to one another, violating one of the main assumptions of this method. The branches seen diverging from the zero line arise because of our choice of using the same random positions for each input model magnitude. The branching occurs when a model galaxy is randomly inserted directly on top of a galaxy in the image, resulting in a measured flux that is systematically too large. This demonstrates the importance of objects being isolated in the detection image.

It is true, however, that these simulated galaxies are most likely fit better than real galaxies because the transformation kernel is a perfect match, which would not be the case in reality. To try and test real galaxies, we compared our photometry with SExtractor's "MAG-AUTO" setting in dual image mode, but only on isolated galaxies (\ie those not flagged as possibly contaminated by light from neighbours). This is a less than ideal comparison, because the size of SExtractor's "MAG-AUTO" aperture is determined by the detection image, and is hence smaller than it should be in the IRAC images, especially in the 5.8 and 8.0\um\ images where the PSF is the largest. However, colors between neighbouring bands should not be strongly affected by this, and so plotted in Figure \ref{fig:testfozzy} are the color differences between our photometry and Sextractor's. Again, we found good agreement.

There are many galaxies that are located in the overlap region in each of the GOODS fields. In this case, we have two photometric measurements for the galaxy,  which we averaged to obtain the final result. The two independent results in the overlap region, however, provide a good estimate of the true amount of uncertainty and scatter in our photometry (see Figure \ref{fig:overlap}).  




\clearpage





\end{document}

%% file: tab1.tex
\begin{deluxetable}{lcccc}
\tablewidth{0pt} 
\tablecaption{\label{schechter.tab}Best-fit Schechter parameters}
\tablehead{
\colhead{Field} &
\colhead{Function} & 
\colhead{$\log(\phi^*)$} &
\colhead{$M^*$} & 
\colhead{$\alpha$} 
}
\startdata
	GOODS North & LF & $-$3.2 $\pm$ 0.6 & $-$24.1 $\pm$ 0.8 & $-$1.5 $\pm$ 1.0\\
	GOODS South & LF & $-$3.0 $\pm$ 0.6 & $-$23.7 $\pm$ 0.7 & $-$1.2 $\pm$ 1.0\\
	Combined & LF & $-$3.1 $\pm$ 0.4 & $-$23.9 $\pm$ 0.7 & $-$1.4 $\pm$ 0.6\\
      \hline
      	GOODS North & MF & $-$3.3 $\pm$ 0.4 & 11.3 $\pm$ 0.2 & $-$1.1 $\pm$ 0.6\\
	GOODS South & MF & $-$3.2 $\pm$ 0.4 & 11.3 $\pm$ 0.3 & $-$1.1 $\pm$ 0.6\\
	Combined & MF & $-$3.3 $\pm$ 0.3 & 11.3 $\pm$ 0.2 & $-$1.1 $\pm$ 0.4\\
\enddata
\end{deluxetable}